\documentclass[conference]{IEEEtran}
\usepackage{amsmath}
\usepackage{amsthm,amssymb,amsmath,bm}
\usepackage{subfigure}
\usepackage{amsfonts}
\usepackage{mathtools} 
\usepackage{amsmath}

\usepackage{epsfig}
\usepackage{amssymb}
\usepackage{amsmath}
\usepackage{cite}
\usepackage[utf8x]{inputenc}
\usepackage{multirow}
\usepackage{rotating}
\usepackage{graphicx}
\usepackage{tabularx}
\usepackage{array}
\usepackage{color,soul}
\usepackage{bm}
\usepackage{tikz}
\usepackage{graphicx,dblfloatfix}
\usepackage{blindtext}

\newtheorem{theorem}{Theorem}

\usepackage{subfigure}
\usepackage{moreverb}
\usepackage{epsfig}
\usepackage{amsmath,amssymb,amsthm,mathrsfs,amsfonts,dsfont}
\usepackage{adjustbox,lipsum}
\usepackage{amsfonts}
\usepackage{epsfig}
\usepackage{amssymb}
\usepackage{amsmath}
\usepackage{amsthm}
\usepackage{subfigure}
\usepackage{multirow}
\usepackage{rotating}
\usepackage{graphicx}
\usepackage{tabularx}
\usepackage{array}
\usepackage{anyfontsize}
\usepackage{color,soul}
\usepackage{graphicx,dblfloatfix}
\usepackage{epstopdf}
\usepackage{blindtext}
\usepackage{amsmath}
\usepackage{amsthm,amssymb,amsmath,bm}
\usepackage{subfigure}
\usepackage{amsfonts}
\usepackage{epsfig}
\usepackage{amssymb}
\usepackage{amsfonts}
\usepackage{cite}
\usepackage{graphicx}
\usepackage{fancyhdr}
\usepackage{subfigure}
\usepackage[subfigure]{tocloft}
\usepackage[font={small}]{caption}
\usepackage{subfigure}
\usepackage{tabularx}
\usepackage{tcolorbox}
\usepackage{cite}
\usepackage[linesnumbered,ruled,vlined]{algorithm2e}
\SetKwInput{KwInput}{Input}
\SetKwInput{KwOutput}{Output}
\usepackage{amsthm,amssymb,amsmath,bm}
\usepackage{graphicx}
\usepackage{fancyhdr}
\usepackage{subfigure}
\usepackage[subfigure]{tocloft}
\usepackage[font={small}]{caption}
\usepackage{subfigure}
\usepackage{tabularx}
\usepackage{cite}
\graphicspath{{Figure_S/}}
\allowdisplaybreaks
\usepackage{url}
\usepackage[colorlinks,bookmarksopen,bookmarksnumbered,citecolor=blue,urlcolor=blue]{hyperref}

\begin{document}
%\vspace{-10mm}
\title{ \huge 
Minimizing AoI in Resource-Constrained Multi-Source Relaying Systems with Stochastic Arrivals
 %Minimizing AoI in Multi-Source Relaying Systems with Stochastic Arrivals Under an Average Resource Constraint
}
%\vspace{-1em}
  \author{\IEEEauthorblockN{Abolfazl Zakeri, Mohammad Moltafet, and Markus Leinonen}
 \IEEEauthorblockA{Centre for Wireless Communications – Radio Technologies\\
 University of Oulu, Finland
 \\
 e-mail: \{abolfazl.zakeri, mohammad.moltafet, markus.leinonen\}@oulu.fi
 }
 \and
 \IEEEauthorblockN{Marian Codreanu}
 \IEEEauthorblockA{Department of Science and Technology\\
 Link\"{o}ping University, Sweden \\
 e-mail: marian.codreanu@liu.se}
% \and
% \IEEEauthorblockN{James Kirk\\ and Montgomery Scott}
% \IEEEauthorblockA{Starfleet Academy\\
% San Francisco, California 96678--2391\\
% Telephone: (800) 555--1212\\
% Fax: (888) 555--1212} 
}	
	\maketitle
%		\title{ \huge  Robust Energy-Efficiency SIC Ordering and  Beamforming Design  in 
%			MISO-NOMA-assisted C-RAN Network: High Degrees of Reused Freedom}
		% Under Imperfect CSI}
% 		\author{Abolfazl Zakeri, Mohammad Moltafet,~Markus Leinonen, and Marian Codreanu
%  		%,~\textit{Senior Member, IEEE}
%  		\thanks{A. Zakeri, M. moltafet, are with 
%  		}}
	% <-this % stops a space
%	\markboth{	IEEE Trans. Veh. Technol}%}%
	%\markboth{IEEE Trans. Commun}%
	%{Submitted paper}
	%\maketitle
	\begin{abstract}
	We consider a multi-source relaying system where the sources independently and randomly generate  status update packets
	%. The packets are sent 
	%as a form of status update packets 
	which 
%	the buffer-aided transmitter sends 
	are sent 
 to the destination  with the aid of a buffer-aided relay through unreliable links.
	%that are scheduled  by a transmitter to sent the destination with the help of a relay in an unreliable environment. 
	%The relay %as an intermediate node is buffer-aided and 
	%can receive and forward simultaneously. %, i.e., can support full-duplex communication.
	We formulate a    stochastic optimization problem aiming to minimize the sum average  age of information (AAoI) of sources under per-slot transmission capacity constraints and a long-run average resource constraint. To solve the problem, we  recast it  as a constrained Markov decision process (CMDP) problem and adopt the Lagrangian method. We analyze the structure of an optimal policy for the resulting MDP problem %which  is in the class of either randomized or mixing policies,
	that possesses a switching-type structure. 
	%--------------------- COMMENTED ACCORDING TO NTO PROOFING
	%Specifically, we  show the existence of the switching-type structure theoretically for the case with error-free links 	and numerically for the general cases.
%	in which communication links are reliable
	%an optimal scheduling  policy %and  
	%that gives a  stationary deterministic policy for the considered problem and
	%We propose an algorithm to obtain  the optimal value of the sum AAoI that establishes a benchmark for any scheduling policy in the considered setup. Moreover, we provide a near-optimal stationary policy, which is easy to implement in practice. 
	% Edited for the camera ready (CR)
	We propose an algorithm that obtains a stationary deterministic near-optimal policy,
	%and a tight lower-bound,
	%the optimal value of the sum AAoI,
	establishing a benchmark for the system.
	%, and provides a deterministic policy.
	%, which is easy to implement in practice. 
	%Numerical results are provided to evaluate the system. 
	% Through the simulation, it is shown  that the  stationary deterministic policy achieves near-optimal performance, and when
%\textcolor{blue}{ 
Simulation results show the effectiveness of  our algorithm compared to benchmark algorithms.
%, and the stationary policy has near-optimal performance.
%, especially, when we more emphasize
%and reveal that the sum 
%Also, %the sum AAoI can be significantly reduced at
%the results reveal optimal policy gives higher propriety to schedule the source with small 
%can be significantly reduced at the expense of resource usage
%on resource consumption. 
 %and is highly affected by the reliability of links.
	%Moreover, when 
	%the average resource constraint is loose and the arrival rates  of the status
%updates are high, the policy behaves like a greedy policy.
	\end{abstract}
%\vspace{-1em}
\section{Introduction}
In many emerging applications  
such as the Internet-of-Things, 
cyber-physical systems, and intelligent transportation systems, the freshness of  status information is crucial \cite{AoI_Mag}. The \textit{age of information} (AoI) has been proposed to characterize the information freshness in status update systems
%from a destination perspective  
\cite{EItam_Young,Roy_2012}. The  AoI is defined as the time elapsed since the latest received status update packet was generated. Recently, the AoI has attracted much interest 
in different areas, e.g., queuing systems \cite{MOhammad_1,Marian_Information} and scheduling problems \cite{Deniz_Relay,hatami,Brancu+2Hop,Mohammad_DC}. 
	 %\textcolor{blue}{ 
	 \\\indent
In some status update systems, there is no direct communication link between the source of information and the intended destination or direct communication is costly.
Deploying an intermediate node, typically called a \textit{relay}, in such systems is indispensable
%, and the existence of such a node 
and has an array of benefits, e.g., saving on power usage of wireless sensors
%in the wireless sensor networks
and improving the transmission success probability. 
%In  such a network design, the communication of two nodes is established by an intermediate node, relay.
% ========== COMMENTED
%For example, %in  vehicular networks, 
%vehicle-to-vehicle (V2V) communication  can potentially be established by another vehicle that relays information among vehicles. 
	Recently, the AoI has been  studied in   relaying systems  in  \cite{MOradi_R,Deniz_Relay,Relay_Nikoss,Shroff,2Hop_TWC,Relay_SA}. 
	The  work \cite{Deniz_Relay} studied the AoI minimization in a multi-source relaying system with the \textit{generate-at-will} model (i.e., possibility of generating a new update in any time),
	  	under a transmission capacity constraint for each link.  They provided an optimal scheduling policy for a setting called the error-prone symmetric IoT network. The authors in \cite{MOradi_R} analyzed the AoI in a discrete-time Markovian system for two different relay settings and analyzed the impact of the relay on the AoI. In \cite{Relay_Nikoss}, the authors analyzed   the average AoI (AAoI)   in a two-way relaying system in which two sources exchange status data, considering the generate-at-will model.
	  	The AoI performance under different policies (e.g., a last-generated-first-served policy) in a general multi-hop  networks with single-source was studied in \cite{Shroff}.
In \cite{2Hop_TWC}, the authors studied the AoI in a single-source energy harvesting relaying system with error-free channels, and 
%where bothe source and relay are energy-harvested nodes.
they designed offline and online age-optimal policies.
%]]]]]]]]]
%\textcolor{blue}{
In \cite{Relay_SA}, the authors  considered a  single-source relaying system with  stochastic arrivals where 
	  	 the source communicates with the destination either through the direct link or via a relay.
	  	They proposed two different relaying protocols and  analyzed the AoI performance.
\\\indent
At the same time,  resource limitations are a main bottleneck in optimizing AoI in  status update systems, especially, in power-limited sensor networks. Only a few works, such as \cite{2Hop_TWC} has incorporated  a resource constraint   in the  relaying system in analysis of the AoI. Moreover, most of the discussed works consider single-source relaying systems, e.g., \cite{2Hop_TWC,MOradi_R,Shroff,Relay_SA}. Accordingly, the AoI in \textit{multi-source resource-constrained} relaying systems has not been widely studied yet.
% 	}
 	%and age-optimal scheduling design is still have  not addressed well in such systems.
 %	\\
 	%We note that, all these works  considered the generate-at-will model, and \cite{Brancu+2Hop,2Hop_TWC} studied single-source relaying systems, and \cite{Deniz_Relay} only considered wireless channel capacity constraints. %However, it may not be possible to adopt generate-at-will model in some systems due to, e.g., limitation on status update generation
% and make optimizing the AoI more challenging.
	  	\\\indent
	  	%-------------- WHAT ARE DOME IN THIS PAPER Contribution--------------
	  %	Motivated by these 
In this paper, we consider a \textit{multi-source} discrete-time relaying system %\footnote{Some times relaying systems can be considered as a two-hop network \cite{Brancu+2Hop}.}
 with \textit{stochastic arrivals}
%(different to generate-at-will mode in \cite{Deniz_Relay,2Hop_TWC,Relay_Nikoss}),
where the sources independently  generate different types
%\footnote{
%Communication scenarios in vehicular networks \cite{Vehicule_Book}, mainly, communication  among vehicles or vehicle to roadside infrastructure communications (i.e.,
%roadside units (RSU)  and base stations), could be examples for the considered communication setup. 
%Moreover, our setting is a general two-hop communication network that supports the cases in which the communication links among nodes are unreliable. 
%}, 
 of status update packets. The packets are delivered to the destination via a buffer-aided transmitter with help of a buffer-aided  relay through \textit{unreliable (error-prone)} links under  transmission capacity constraint for each link.
 %, i.e., at each slot, only one source can be scheduled per link. 
 We further consider a \textit{long-run average resource constraint} on the average number of all transmissions (i.e., transmitter-to-relay and relay-to-destination). %, which represents e.g., the total power consumption. 
 %different to \cite{2Hop_TWC, Deniz_Relay}. 
 %The relay can receive and forward simultaneously. 
 %, i.e., it works in the \textit{full-duplex} operation mode. 
 The considered  setup can be a representative of,
 % communication scenarios,
  e.g.,  %in vehicular networks \cite{Vehicule_Book}, mainly,
vehicle-to-vehicle (V2V) communications,  and vehicle-to-infrastructure (V2I) communications in which the infrastructure could be 
roadside units (RSUs)  or base stations and the communications are established with the help of another vehicle acting as a relay  \cite{Vehicule_Book}.
%Our aim is  designing 
We focus to design an \textit{age-optimal scheduling policy} with the sum  AAoI metric
%that  minimizes the sum average  AoI (AAoI) of sources 
under the  transmission capacity and average resource constraints. 
%In specific,   for each link (transmitter-relay and relay-destination) a per-slot  transmission capacity constraint, i.e., only one source can be scheduled, is considered. 
%To provide a practical system and real-world analysis,
 %To design an age-optimal scheduling policy, 
Specifically, we formulate a stochastic optimization problem and then recast it as a constrained Markov decision process (CMDP) problem.
By  adopting the Lagrangian method, we analyze the structure of an optimal policy and propose an algorithm that finds a stationary deterministic near-optimal policy. %, establishing a benchmark in the considered system.
%================ COMMENTED
%We  show that an optimal policy has a switching-type structure theoretically for the case with error-free links and empirically  for the general case.
%========================
%\textcolor{blue}{
%Moreover, we provide a near-optimal stationary policy, which is easy to implement  in practice. 
Simulation results show the effectiveness of  our algorithm compared to benchmark algorithms. %}
%, especially, when the allowable average number of transmissions is small.}
%}
%Simulation results demonstrate the effectiveness of our proposed algorithm and  show  that the sum AAoI is highly affected by the availability of resources and the reliability of transmissions.
%in which the links are unreliable.
	  	\\\indent
	  	%--------------------------- Most Related WORKS------------------
	 % \textcolor{blue}{
	 Our relaying system as a \textit{two-hop} network is an extension of the recent work
%a more general and important  extension of the  recent salient work 
\cite{Eyton_Modiano}, where the authors provided scheduling algorithms for minimizing AoI in a \textit{one-hop buffer-free} network with stochastic arrivals and  an \textit{error-free} link,  with no resource constraint. %under the transmission capacity constraint.
In contrast, in our two-hop buffer-aided network, all communication links are error-prone, and 
we consider an average resource constraint.
%and the transmission capacity constraint per link.
	  The most related work to our paper is  \cite{Brancu+2Hop}, where 
	  	% The  work \cite{Deniz_Relay} studied the AoI minimization in a multi-source relaying system with the \textit{generate-at-will} model,
	  	% under a transmission capacity constraint for each link.  They provided an optimal scheduling policy for a setting called the error-prone symmetric IoT network.
	  %	A recent work \cite{Brancu+2Hop}
	  	the authors studied  the AoI minimization problem in a single-source relaying system with  the generate-at-will model under a resource constraint  on the average number of forwarding transmissions at the relay. 
       Different to \cite{Brancu+2Hop},  we consider a multi-source setup with stochastic  arrivals  which is a challenging generalization of the
       %single-source {generate-at-will} 
       model adopted in \cite{Brancu+2Hop}. In addition,
       %. In addition, 	%different to \cite{Deniz_Relay}, we consider an average resource constraint %(besides the transmission capacity constraints) 
      % and  we provide  a structural optimal policy by using the CMDP analysis;
      % different to \cite{Brancu+2Hop}, 
     %  and  a  multi-source setup and 
       we consider a global resource constraint on the average  number of  all transmissions in the system.
	\section{System Model and Problem Formulation}
	\subsection{System Model}
\begin{figure}%[h!]
    \centering
    \includegraphics[width=.45\textwidth]{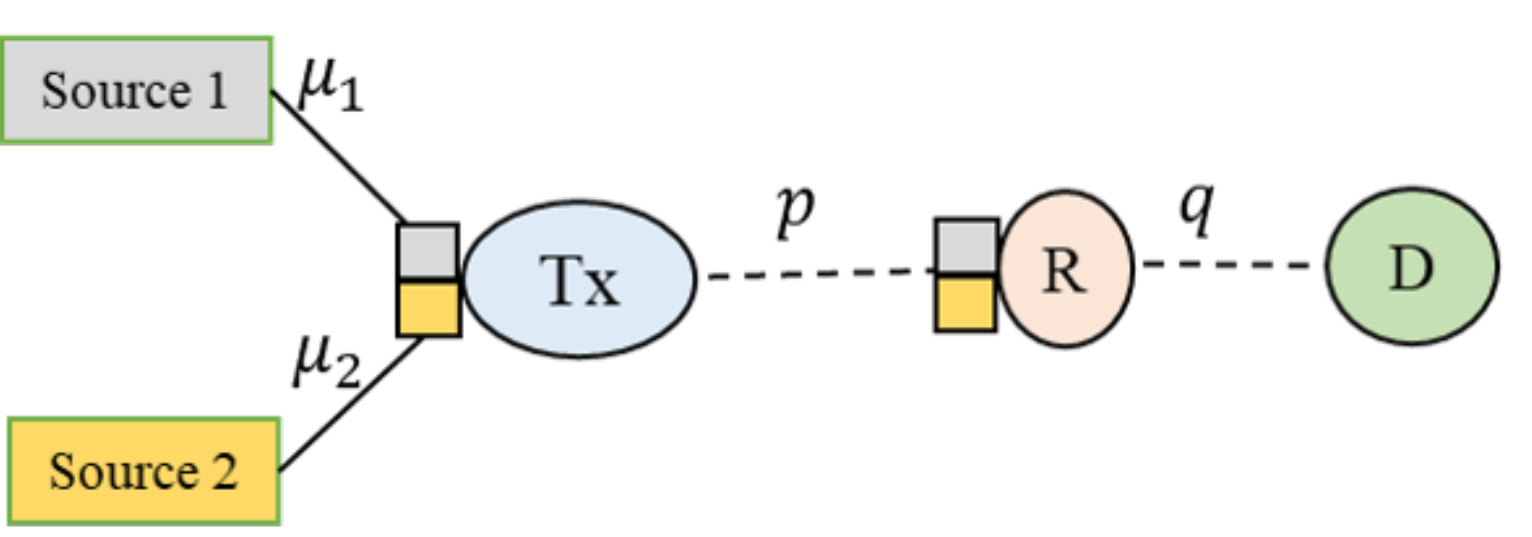}\vspace{-1mm}
    \caption{A multi-source relaying status update system in which different status updates arrive at random time slots at Tx which then sends the packets to D via R. For example, in a vehicular setting,  sources can be considered as different status information (e.g., speed and location), Tx can be one vehicle's transmitter  called on-board unit \cite{Vehicule_Book}, R   can be another vehicle, and D can be an RSU or a vehicle.
    \vspace{-2em}
    }
    %and 2) sources can be considered as a different  information (e.g., traffic and transportation
%information for planning the best route \cite{Eyton_Modiano}), Tx can be a macro base station, R   can be a small base station, and D can be an end-user.}
    \label{SM}
\end{figure}

	We consider a status update system  consisting of two independent sources\footnote{We consider two sources for simplicity of presentation. The formulations and analyses can be  extended for more than two sources. However, the complexity of computation, especially in numerical analysis, increases exponentially  with the number of sources.}, a buffer-aided transmitter (Tx), a buffer-aided\footnote{Note that the buffers at Tx and R enable a re-transmission mechanism that can improve the AoI, in particular, when the arrival rate and transmission success probability are small. We assume that all re-transmissions have the same transmission success probability.} relay (R), and a destination (D), as depicted in  Fig. \ref{SM}.
 We assume that % there is no direct link between Tx and D. Thus,
Tx sends status update packets to D via  R
and there is no direct communication between Tx and D. Further, we assume that the size of the buffer in Tx and R is  one packet per source. We assume that each transmission takes one slot duration.
%	The relay  can receive and forward  simultaneously.
	\\\indent
	Time is slotted and $t=0,1,2,3,\dots$, indicates the slot index. 
The sources, indexed\footnote{Herein, $i$ denotes the index of sources and it always varies from $1$ to $2$.} by $i\in\{1,2\}$,   generate status update packets independently according to the Bernoulli  distribution with parameter $\mu_i$, where the packets arrive at Tx at the beginning of slots. %\textcolor{blue}{
We assume that the old packets in the waiting buffers %, and a new one arrives from the same source, Tx (resp. R)
are replaced with the new ones from the same source, which is done at the beginning of each slot.
%}
Let $\Lambda_i[t]$ be a binary indicator that shows whether a packet from  source  $i$  arrives at Tx  in slot $t$, where $\Lambda_i[t]=1$ denotes that a packet arrived; otherwise, $\Lambda_i[t]=0$. Thus, $\Pr\{\Lambda_i[t]=1\}=\mu_i$. 
%, where	$\Pr\{.\}$ denotes the probability function.  
%As explained before, multi-source scenarios also interested. 
%The main notations are listed in Table \ref{Table_Not}.
% 	\begin{table}
% 				\caption{Main notations}
% 		\label{Table_Not}
% 		\centering
% 		\renewcommand{\arraystretch}{1.3}
% 	\begin{tabular}{c||l}
% 		\hline\textbf{Notation(s)} & \textbf{Definition} \\
% 	%	\hline  \mu_ilticolumn{2}{l}{~~~~~~~~~~~~\textbf{Notations/parameters}}
% 	%	\\
% 		\hline$\beta_t$ & Sampling variable \\
% 		\hline$\alpha[t]$ & Transmission variable \\
% 		\hline $\Delta_t$ & Age of information\\
% 		\hline $\Delta_t^{\text{Th}}$ & Denotes the threshold variable \\
% 		\hline
% 		$X_t$& Value of the process in the source     side at time $t$
% 		\\
% 		\hline
% 		$\hat{X}_t$& Value of the process in the monitor side at time $t$
% 		\\\hline
% 		$ C_t^{\text{Tot.}}$ & Total transmission and sampling cost
% % 		\hline \mu_ilticolumn{2}{l}{~~~~~~~~~~~~\textbf{ Optimization Variables}} %($ t $ $\rightarrow$ denotes time slot )
% % 		\\
% % 		\hline$\varphi_{m}^{f}[t]$ & Binary transmission indicator   for information \\& $ f $ of UE $ m $ \\
% % 		\hline$\xi_{m,n}[t]$ & Binary subcarrier assignment of UE $ m $ \\&on subcarrier $ n $
% % 		\\ \hline$p_{m,n}[t]$ &Transmit power   of UE $ m $ on subcarrier $ n $
% % 		\\
% % 		\hline
% 	\end{tabular}
% 	\end{table}
 \\
 $ \bullet$
  \textbf{Wireless Channels:}
 We assume an unreliable channel with transmission success probability $p$ and $q$ for the  Tx-R link and  R-D link, respectively. { Let $\rho [t]$ (resp. $\rho' [t]$) be a binary indicator that represents the  success of a packet transmission  at the   Tx-R link (resp. the  R-D link) in slot $t$.} In specific,
 if $\rho [t]=1$ (resp. $\rho'[t]=1$), the  transmitted packet in slot $t$ from  Tx (resp. R) is successfully received by R (resp. D); otherwise, $\rho [t]=0$ (resp. $\rho'[t]=0$). 
% Similarly, if $q[t]=1$, the  transmitted packet from  R is successfully received by  D, otherwise, $q[t]=0$.
Thus, $\Pr\{\rho [t]=1\}=p$ and  $\Pr\{\rho'[t]=1\}=q$.
We assume that the perfect feedback (delay-free and error-free) is available at each link. 
  \\$ \bullet$ \textbf{Decision Variables:}
  We assume that, in each slot, at most one  source can be scheduled per link and the transmissions are over orthogonal channels.
% in both the Tx-R link and the R-D link.}
  %, i.e., the system imposes per-slot hard transmission capacity constraints.
 %Thus, at each link, at most one source  can be scheduled in each slot.
 Let $\alpha[t]\in\{0,1,2\}$  denote the (transmission) decision of  Tx in slot $t$, where $\alpha[t]=i$ %for $i=1,2$ 
 means that Tx transmits the packet of source $i$ to R, and $\alpha[t]=0$ means Tx stays idle.
 Similarly, let  $\beta[t]\in\{0,1,2\}$ denote the (transmission) decision of  R in slot $t$, where $\beta[t]=i$  %$i=1,2$ 
 means that R forwards the packet of source $i$ to D, and  $\beta[t]=0$ means that R stays idle. 
 %By such definition of the decision variables, we  take into account the per-slot  hard transmission constraints (due to hard limits on  radio  resources, e.g., bandwidth)  of the links Tx-R and R-D. Thus, at each link, at most one source  can be scheduled in each slot.
 %\footnote{Such definition reduces the action space (will be introduced) and simplifies formulation, instead of, using binary scheduling variables for each source and corresponding constraint.} 
  We assume that there is a centralized controller that decides what Tx and R does during each slot.  
 %that Tx is the central controller %\footnote{In this paper, we focus to devise a centralized decision-maker.}
 %Distributed algorithm design is deferred to future works.} 
 %and %it knows about  the AoIs. Indeed, it 
 %informs the related decision ($\beta[t]$) to R with negligible delay at the beginning of each slot \cite{Relay_SA}. 
%\\  $ \bullet$ \textbf{Packet Management:}
%Moreover, when Tx  (resp. R)  has a 
 %Moreover, if Tx (resp. R) receives the ACK message from R (resp. D), Tx (resp. R) discards the previously sent packet. 
%Also, all re-transmitted packets have the same loss probability. % similar to \cite{Deniz_Relay}. %\cite{Deniz_Relay}
 % coming from the same source.}
 %``idle" and  $\beta[t]=1$ denotes ``transmission".
 %, and $\beta[t]=2$ denotes ``transmitting packet of source 2", respectively.
 %------------ Low action space
% \textbf{Packet managemnet}
%==============================
%  \begin{align}
%  \begin{split}
%   \alpha[t] \in \Bold{A}_{t}^{\text{Tran.}}= \begin{cases}
%         (i,t) & K_t=1,\beta_t=1,
%         \\
%          (i,t,r) & K_t=0,\beta_t=1,
%          \\
%           (i,t) & K_t=1,\beta_t=0,
%           \\
%           (i,r) & K_t=0,\beta_t=0,
%       \end{cases}
%  \end{split}
%  \end{align}
%  where $\Bold{A}_{t}^{\text{Tran.}}$ denotes all possible actions for $\alpha[t]$.
%=====================
\\$\bullet$ \textbf{AoI:} Let $\theta_i[t]$, $\delta_i[t]$, and  $\Delta_i[t]$ be the AoI  of source $i$
at Tx, R, and D, in slot $t$, respectively. The evolution of these AoIs are given by
%on  top of the next page.
%\\
%\begin{figure*}
\begin{align}
%\scriptsize
%\footnotesize
%\small
%\[
%\label{AoI_R1}
&\nonumber
    \theta_i[t+1]=\begin{cases}%\begin{array}{ll}
     0,  & \text{if}~~ \Lambda_i [t+1]=1, \\
        \theta_i[t]+1, & \text{otherwise}
 %  \thet\bold{a}[t]\mathds{1}_{\{\alpha[t]=1\}}+
 %  \theta_i[t]
   % \end{array}
    \end{cases}, %~i\in\{1,2\},
%\end{align}
%The evolution of AoI for 
%$i\in\mathcal{I}$ 
%at the R can be characterized by:
%\begin{align}
%\\&
\\&
\nonumber
%\label{AoI_R1}
    \delta_i[t+1]=\begin{cases}
 %  \thet\bold{a}[t]\mathds{1}_{\{\alpha[t]=1\}}+
   \theta_i[t]+1 &\text{if}~~ \alpha[t]=i,~\rho [t]=1,
    \\
    \delta_i[t]+1, & \text{otherwise}
    \end{cases}, %~i\in\{1,2\},
%\end{align}
%The evolution of AoI at the destination can be characterized by
%\begin{align}
\\\nonumber
&
    \Delta_i[t+1]=\begin{cases}\label{AoI_D1}
    \delta_i[t]+1, &\text{if}~~
    \beta[t]=i,~\rho'[t]=1,    %    \delta_{i}[t]+1, &\text{if}~~ q[t]=1,~\beta[t]=i,
    \\
    \Delta_i[t]+1, &\text{otherwise}
    \end{cases}. %~i\in\{1,2\}.
   % \]
\end{align}
%\hrule
%\vspace{-1em}
%\end{figure*}
% \begin{Asu}
% (Assumption on the Controller): We assume, Tx is central controller, and %it knows about  the AoIs. Indeed, it 
% informs the related decisions ($\beta[t]$) to R with negligible delay at the begging of each slot.
% \end{Asu}
%\textbf{Performance Metric and Constrain:} In our setting, we interested to 
%\vspace{-5mm}
\subsection{Problem Formulation}
% We are intent to minimize long-term average AoI under long-term transmission and sampling cost by optimizing the sequence of decision variables.
 %\\
% \textcolor{blue}{
 Let %$\pi$
 $\mathcal{D}=\{(\alpha[t],\beta[t])~\big|~\alpha[t],\beta[t]\in\{0,1,2\}\}$, $t=1,2,3,\dots$, be a sequence of decision variables.
% be a scheduling policy that determines the decision variables of Tx and R in each slot $t$, i.e.,  is a rule that generates these decision variables. 
 We intend to optimize the decision variables in order to minimize the time average sum of AoIs at D,  satisfying: i) %per-slot constraint
% at each slot only one source can be scheduled per link, %as a 
 the transmission capacity constraint per link at every  slot
% per-slot hard transmission capacity constraints for the Tx-R and R-D links  
 and ii) a long-run average resource constraint.
 Let $\bar{\Delta}(\mathcal{D})$ and $\bar{c}(\mathcal{D})$ denote the long-run expected (time) average  sum of AoIs at D and the average number of transmissions in the system, respectively, for given  $\mathcal{D}$, which are defined as 
 %}
  \begin{equation}
     \begin{array}{ll}
     \nonumber 
    & \displaystyle \bar{\Delta}(\mathcal{D}) \triangleq\limsup_{T\rightarrow \infty} \frac{1}{T}
        \Bbb{E}\left[\sum_{t=1}^{T} %\left(
        \Delta_1[t]+\Delta_2[t] %+\Delta_2[t]\right)
        \right],
        \\&\nonumber
        \displaystyle\bar{c}(\mathcal{D})\triangleq\limsup_{T\rightarrow \infty} \frac{1}{T} \Bbb{E}\left[\sum_{t=1}^{T} 
        \mathds{1}_{\{\alpha[t]\neq0\}}+\mathds{1}_{\{\beta[t]\neq 0\}}\right],
      \end{array}
 \end{equation}
 \iffalse
 \begin{align}
     \nonumber 
    & \bar{\Delta}(\pi) \triangleq\limsup_{T\rightarrow \infty} \frac{1}{T}
        \Bbb{E}\left[\sum_{t=1}^{T} %\left(
        \Delta_1[t]+\Delta_2[t] %+\Delta_2[t]\right)
        \right],
        \\&\nonumber
        \bar{c}(\pi)\triangleq\lim_{T\rightarrow \infty} \sup\frac{1}{T} \Bbb{E}\left[\sum_{t=1}^{T} 
       \left(
       \mathds{1}_{\{\alpha[t]\neq0\}}+\mathds{1}_{\{\beta[t]\neq 0\}}\right)
       %}{}\big|s[0]
       \right],
      % \le \Gamma_{\max},
 \end{align}
 \fi
    where %$i\in\{1,2\}$ and
   $\mathds{1}_{\{\cdot\}}$ is an indicator function  which   equals to $1$ when the condition in $\{\cdot\}$ holds, and
   $\Bbb{E}\{\cdot\}$ is the expectation with respect to the system randomness (i.e., channel reliability and packet arrival processes)
   and the policy.
By these definitions,  our aim is to solve the following stochastic optimization problem
\vspace{1em}
%by optimizing scheduling policy $\pi$:
        \begin{subequations}\label{Org_P1}
   \begin{align}
      % \mathcal{OP}1:~~ 
  %     \\
  %\nonumber
     \underset{\mathcal{D}}{\text{minimize}}~~~~
        %\boldsymbol{\alpha}[t],\boldsymbol{\beta}[t]}
          %J(\pi)
        %\bar {\Delta}(\pi)
        %\triangleq
      &  \bar{\Delta}(\mathcal{D})
       %\triangleq \limsup_{T\rightarrow \infty} \frac{1}{T}
      %  \Bbb{E}\left[\sum_{t=1}^{T} %\left(
      %  \Delta_1[t]+\Delta_2[t] %+\Delta_2[t]\right)
       % \right],
        \\%\nonumber
      \text{subject~to}~~~~ &
      %\label{Con_Tra1}
      \bar{c}(\mathcal{D})
       \le \Gamma_{\max}, 
    \label{Con_Sam1}
            \end{align}
     \end{subequations}
 where the real value   $\Gamma_{\max}\in(0,2]$ is the maximum allowable average number of transmissions in the system. Constraint \eqref{Con_Sam1} is an average resource  constraint that reflects  the resource limitation in the normalized form, i.e., normalized to the power usage per each transmission. %\textcolor{blue}{
 Thus the constraint can represent a limitation on the total average consumed power. %}
 \vspace{-1mm}
\section{ CMDP Formulation and Lagrange Relaxation }
%\textit{{Problem 1}} %\eqref{Org_P1}
%is in the class of the stochastic optimization problem \textcolor{red}{Make sure or change} with infinite horizon
%sate space (due to the growth of AoI) and possibly unbounded per-slot cost function
%with expected average cost criteria which is generally complicated to find the optimal policy.   However,
%next we provide some  theoretical results that give the structure of the optimal policy which are useful to design a practical algorithm to derive the policy. 
In this section, we attain to solve main problem \eqref{Org_P1} by transforming it into 
%Before going to the details, we present the main items and results incorporated in this section.
%\subsection{Main Results}
%\begin{itemize}
        %\item 
%To this end, first, we transform it into 
a CMDP
%\footnote{It is noteworthy that according to  the system states, transition to the next states of the system only depends on the current state, taken action, and randomness of the system (see Eq. %s \eqref{AoI_R1}-\eqref{AoI_D1} and
%\eqref{Eq_TranPro_Unr}). Therefore, the considered system has the Markov property. Moreover, the model is stationary.} 
problem which is then solved by
%and then 
using the Lagrangian relaxation method.
\subsection{CMDP Formulation}
%We recast problem \eqref{Org_P1} as a
We introduce the CMDP by the following elements:
%where $\mathcal{S}_N $ denotes the set of sates, $\Bold{A}$ is the set of actions, $c$ is the immediate cost function, D  is the immediate constraint cost for transmission, $e$ is the immediate constraint cost for sampling, and
%$\mathcal{P}$  is the transition probability distribution, and $s_0$ is the initial state.
%\begin{itemize}
    %\item 
    \\$\bullet $
    \textbf{State}:
    %\textcolor{blue}{ 
  The  state of the CMDP  incorporates the knowledge about the AoIs at Tx, R, and D.
    We define the state in slot $t$ by
 $ \bold{s}[t]\triangleq (\theta_1[t],x_1[t],y_1[t],\theta_2[t],x_2[t],y_2[t])$,
 where $x_i[t]\triangleq \delta_i[t]-\theta_i[t]$ and $y_i[t]\triangleq \Delta_i[t]-\delta_i[t]$ are the \textit{relative AoIs} at R and D in slot $t$, respectively. Using the relative AoIs simplifies 
 %we can define the AoI at D in each slot, only as a function of state elements, which 
 the subsequent analysis and derivations. 
 The intuition is that the amount of
 the change of the sum AoIs at  D from slot $t$ to the next slot $t+1$  is determined by $y_i[t]$,
 %is affected by the decision $\beta[t]$ and the amount of that change is determined by $y_i[t]$, 
 which is shown in the following equation
 %than including them individually
    %\\
    %\\
   % However, instead of including them individually, we incorporate \textit{relative AoIs} in the state, which
   % simplify the subsequent analysis and derivations.
    %of the structure of an optimal policy and  the equation of  sum AoIs at D.
  %  the objective function 
   % which  is defined (in the next) as
 %   an operating cost in the system.
   % To this end, we rewrite the sum AoIs at D
    %Typically, the state of our considered  system could be the AoIs at Tx, R, and D \cite{Deniz_Relay,Eyton_Modiano}.
    %Meanwhile, some insight can be gained by rewriting 
    %the total  AoI at D, 
   % in slot $t+1$ as 
    \[
    \sum_i\Delta_i[t+1]-\sum_i \Delta_i[t]=2-\mathds{1}_{\{\beta[t]=i,\rho'[t]=1\}}(\underbrace{\Delta_i[t]-\delta_i[t]}_{y_i[t]}).
    \]
    Similarly, the amount of the change of the AoI at R is
   % where $\delta_i[t]$ is affected by decision on
 %  $\alpha[t-1]$ with amount
   $x_i[t]$. 
    {Moreover, we denote the state space by  $\mathcal{S}$ which includes all possible states. Note that $\mathcal{S}$ is a \textit{countable infinite} set due to the AoIs being potentially unbounded.
   % , possibly with infinite cardinality.
    }
     \\$\bullet $ \textbf{Action}:   
   %  \textcolor{blue}{
     We define 
     the action   taken in slot $t$ by $\bold{a}[t]=(\alpha[t],\beta[t])$, where $\alpha[t],\beta[t]\in\{0,1,2\}$.
     %, e.g., $\alpha[t]=1$ means that Tx sends a packet of source $1$ to R in slot $t$.  
    % The actions are the decision variables in the system, %which 
     Actions are determined by a policy, denoted by $\pi$, which  is a rule that generates
these actions by observing the current state (i.e., Markovian policies), i.e., a policy is a mapping from states to actions, potentially with a probability distribution.
     % Let  be a scheduling policy that determines the decision
%variables of Tx and R in each slot t, i.e., is a rule that generates
%these decision variables
    %, respectively, for the Tx side and the relay side.
    Moreover, let $\mathcal{A}$
  denote the action space. 
  %}
    %--------------------
   % that includes all feasible actions at state $\bold{s}$, which is finite with cardinality $|\mathcal{A}_{\bold{s}}|\le9$. The feasible actions at each state are determined by taking into account the following rule. At each state in which $x_i=0$ (resp. $y_i=0$), the action cannot be $\alpha=i$ (resp. $\beta=i$) because $x_i=0$ (resp. $y_i=0$) represents the situation that there is no packet of source $i$ at the buffer of Tx (resp. R). More precisely, $x_i=0$ (resp. $y_i=0$) implies that R and Tx (resp. D and R) possess the same level of freshness with respect to status update of source $i$.
    %------------------------
   % }
   % Each action in $\mathcal{A}$ is denoted by $\bold{a}=(\alpha,\beta)$.
  %  Actions includes 
      \\$\bullet $ \textbf{Cost Functions}:
    %  \textcolor{blue}{
      The (immediate) cost functions include: 1) the AoI cost, and 2) the transmission cost.
    The  AoI cost  of each  slot $t$ is the sum AoIs at D %, and %, defined at the beginning of each slot.
    %By  taking an action $\bold{a}[t]$, the change of AoI is reflected at the beginning of next slot $t+1$. 
   % For given $\bold{s}[t]$ and action $\bold{a}[t]$
   % The AoI cost 
 %   for  slot $t$ is 
    given by
%  \begin{align}\nonumber
%  %\[
% $
%$
%\[
$C(\bold{s}[t])=\sum_i \theta_i[t]+x_i[t]+y_i[t].$
%\]
%\end{align}
The transmission  cost  of  slot $t$, for action $\bold{a}[t]$ is %(does not a function of states) 
 defined by
% \begin{equation}
% \nonumber
%  \begin{array}
  %\[ 
  $D(\bold{a}[t])=\mathds{1}_{\{\alpha[t]\neq 0\}}+\mathds{1}_{\{\beta[t]\neq 0\}}.$
 % \]
  %}
%  \end{array}
% \end{equation}
%ADD IN THE NEXT \|\bold{A}[t]\|_0
%\mathds{1}_{\{\alpha[t]= i\}}+\mathds{1}_{\{\beta[t]\neq 0\}}, a=(\alpha[t],\beta[t])\in\Bold{A}.$$
%$$ d(a)=
%\mathds{1}_{\{\alpha\neq 0\}}+\mathds{1}_{\{\beta\neq 0\}}, a=(\alpha,\beta)\in\Bold{A}.$$
%$   
%\end{align}
%3) Sampling cost  $e: \Bold{A}\rightarrow \Bbb{R}_{+}$, is given by 
% \begin{align}
%     e[t]=\mathds{1}_{\{\alpha[t]= 2\}}. 
% \end{align}
%\end{align}
%\\
     \\$\bullet $ \textbf{State Transition Probabilities}: For any two states $\bold{s},\bold{s}'\in\mathcal{S}$, $\mathcal{P}_{\bold{s}\bold{s}'}(\bold{a})$ is the state transition probability
     %\footnote{Here, $t$ is omitted, because, in each transition, the current state and action are important, not which slot is.}
     that gives the probability of moving to state $\bold{s}'$ (next state) from state $\bold{s}$ (current state) under taking an action $\bold{a}=(\alpha,\beta)$.
  Mathematically,   
  $\mathcal{P}_{\bold{s}\bold{s}'}(\bold{a})$ is given as
  $ %\begin{align}\nonumber
  %\label{Eq_TransPro}
     \mathcal{P}_{\bold{s}\bold{s}'}(\bold{a})=\prod_{i}   %\in\mathcal{I}
     %\prod_{i:\Lambda'_i=1} \mu_i  \prod_{i:\Lambda'_i=0} (1-\mu_i) 
     \Pr\{\bold{s}_i' 
     %_{\Lambda'}=(\delta',x')
     |\bold{s}_i,\bold{a}\}$, 
  %\end{align}
  %\vspace{-2em}
%     \begin{align}
%      \Pr\{\bold{s}_i'|\bold{s}_i,\bold{a}\} =
%      \prod_{\alpha}\prod_{\beta}
%      %\prod_{i:\Lambda'_i=1} \mu_i  \prod_{i:\Lambda'_i=0} (1-\mu_i) 
%      \Pr\{\bold{s}_i' 
%      %_{\Lambda'}=(\delta',x')
%      |\bold{s}_i,\bold{a}\},~~\bold{s}'_i=(x'_i,y'_i),\bold{s}_i=(x_i,y_i),
%   \end{align}
%\textcolor{blue}{
  where $\bold{s}'_i=(\theta'_i,x'_i,y'_i),~\bold{s}_i=(\theta_i,x_i,y_i)$
  are state vectors associated to source $i$, 
 % }
  and $\Pr\{\bold{s}_i'|\bold{s}_i,\bold{a}\}$ is given by 
  %===========CMNETE
   \eqref{Eq_TranPro_Unr} shown on the top of next page. Note that $\bold{s}$ is arranged as $\bold{s}=(\bold{s}_1,\bold{s}_2)$.
    \begin{figure*}
%     %% Transition of Unreliable Setting 
%\begin{s
%\\
       % \begin{subequations} 
        \begin{align}
    %\tiny
    \small
   %\scriptsize
  %\fontsize
     \Pr\{\bold{s}_i'|\bold{s}_i=(\theta_i,x_i,y_i),\bold{a}=(\alpha,\beta)\}= \label{Eq_TranPro_Unr}
    %\tag{\ref{Eq_TranPro_Unr}}%\notag\\
      %\Pi_{\Lambda'=1}\mu_i\Pi_{\Lambda'=0}(1-\mu_i)
   % \nonumber
    %\\
    \begin{cases}
      %\begin{array}{ll}
      \mu_ipq,& \alpha=i,\beta=i;~ \theta'_i=0,~x'_i=\theta_i+1,~y'_i=x_i,
      \\
      \mu_i(1-p)q,& \alpha=i,\beta=i;~ \theta'_i=0,~x'_i=x_i+\theta_i+1,~y'_i=0,
      \\
      \mu_ip(1-q), & \alpha=i,\beta=i;~ \theta'_i=0,~x'_i=\theta_i+1,~y'_i=y_i+x_i,
      \\\mu_i(1-p)(1-q), & \alpha=i,\beta=i;~ \theta'_i=0,~x'_i=x_i+\theta_i+1,~y'_i=y_i,
      \\  (1-\mu_i)pq, & \alpha=i,\beta=i;~ \theta'_i=\theta_i+1,~x'_i=0,~y'_i=x_i,
      \\
      (1-\mu_i)(1-p)q, & \alpha=i,\beta=i;~ \theta'_i=\theta_i+1,~x'_i=x_i,~y'_i=0,
      \\
      (1-\mu_i)p(1-q), & \alpha=i,\beta=i;~ \theta'_i=\theta_i+1,~x'_i=0,~y'_i=y_i+x_i,
      \\
    {(1-p)(1-q)(1-\mu_i)}, & \alpha=i,\beta=i;~ \theta'_i=\theta_i+1,~x'_i=x_i,~y'_i=y_i,
      \\
          \mu_ip, & \alpha=i,\beta\neq i;~ \theta'_i=0,~x'_i=\theta_i+1,~y'_i=y_i+x_i,
          \\
          \mu_i(1-p), & \alpha=i,\beta\neq i;~ \theta'_i=0,~x'_i=x_i+\theta_i+1,~y'_i=y_i,
      \\  (1-\mu_i)p, & \alpha=i,\beta\neq i;~ \theta'_i=\theta_i+1,~x'_i=0,~y'_i=y_i+x_i,
      \\
      (1-\mu_i)(1-p), & \alpha=i,\beta\neq i;~ \theta'_i=\theta_i+1,~x'_i=x_i,~y'_i=y_i,
  \\
          \mu_iq, & \alpha\neq i,\beta= i;~ \theta'_i=0,~x'_i=x_i+\theta_i+1,~y'_i=0,
          \\
          \mu_i(1-q), & \alpha\neq i,\beta= i;~ \theta'_i=0,~x'_i=x_i+\theta_i+1,~y'_i=y_i,
      \\  (1-\mu_i)q,& \alpha\neq i,\beta= i;~ \theta'_i=\theta_i+1,~x'_i=x_i,~y'_i=0,
      \\
      (1-\mu_i)(1-q),& \alpha\neq i,\beta= i;~ \theta'_i=\theta_i+1,~x'_i=x_i,~y'_i=y_i,
\\           \mu_i, & \alpha\neq i,\beta\neq i;~ \theta'_i=0,~x'_i=x_i+\theta_i+1,~y'_i=y_i,
      \\  1-\mu_i,& \alpha\neq i,\beta\neq i;~ \theta'_i=\theta_i+1,~x'_i=x_i,~y'_i=y_i,
      \\ 0& \text{otherwise},
     % \end{array}.
      \end{cases}
      \end{align}
   % \end{subequations}
   \hrule
   \end{figure*}
   \\\indent
By these definitions,  the objective of CMDP,
%average operating cost of the CMDP, 
denoted by $J(\pi;\bold{s}[0])$, which is the expected average sum AoIs at D, is  given by
%     \begin{align}
  %   \nonumber
          % & 
           %$
           \[
           J(\pi;\bold{s}[0])=
        \limsup_{T\rightarrow \infty} \frac{1}{T}
        \Bbb{E}\left[\sum_{t=1}^{T} %\left(w_1\Delta_1[t]+w_2\Delta_2[t]\right)
        C(\bold{s}[t])~
        %\sum_{i=1}^{2} w_i\Delta_i[t]
        \big|~\bold{s}[0]
        \right],\]  %$,
    %     \\&
    %     \bar{D}(\pi,\bold{s}[0])=\limsup_{T\rightarrow \infty} \frac{1}{T} \Bbb{E}\left[\sum_{t=1}^{T} {
    %   D(\bold{a}[t])
    %   %\left(\mathds{1}_{\{\alpha[t]\neq 0\}}+\mathds{1}_{\{\beta[t]\neq 0\}}\right)
    %   }{}\big|\bold{s}[0]
    %   \right]
       %\le \Gamma_{\max},
     %\end{align}
     where $\bold{s}[0]$ is the initial state.
     Similarly,  the constraint function of %the long-run average holding cost of 
     CMDP, denoted by $\bar{D}(\pi;\bold{s}[0])$, which is the expected average transmission cost in the system, is given by
    % \begin{equation}
    %    \nonumber
    \[
    \bar{D}(\pi;\bold{s}[0])=
   % \limsup{dd}
    \limsup_{T\rightarrow \infty} \frac{1}{T} \Bbb{E}\left[\sum_{t=1}^{T} {
       D(\bold{a}[t])
       %\left(\mathds{1}_{\{\alpha[t]\neq 0\}}+\mathds{1}_{\{\beta[t]\neq 0\}}\right)
       }{}~\big|~\bold{s}[0]
       \right].\] Now, problem \eqref{Org_P1} is transformed into  the following CMDP problem 
       %\vspace{-1em}
      \begin{subequations}\label{Org_P}
   \begin{align}
       %\mathcal{OP}2:~~ 
      % \nonumber
         \underset{\pi}{\text{minimize}}~~~~ &
        %\boldsymbol{\alpha}[t],\boldsymbol{\beta}[t]}
        J(\pi;\bold{s}[0])
        %\triangleq
       % \limsup_{T\rightarrow \infty} \frac{1}{T}
       % \Bbb{E}\left[\sum_{t=1}^{T} %\left(w_1\Delta_1[t]+w_2\Delta_2[t]\right)
       % C(\bold{s}[t])
        %\sum_{i=1}^{2} w_i\Delta_i[t]
       % \big|\bold{s}[0]
       % \right],
        \\%\nonumber
       \text{subject~to}~~~~ &\label{Con_Tra1} \bar{D}(\pi;\bold{s}[0])
    %   =\limsup_{T\rightarrow \infty} \frac{1}{T} \Bbb{E}\left[\sum_{t=1}^{T} {
    %   D(\bold{a}[t])
    %   }{}\big|\bold{s}[0]
    %   \right]
       \le \Gamma_{\max}.
           \end{align}
          % \vspace{-0.5em}
     \end{subequations}
%\end{Problem}
The optimal value of  problem \eqref{Org_P} for a given $\bold{s}[0]$ is denoted by $J^*(\bold{s}[0])$.
%Next, we attain to solve problem \eqref{Org_P} using the Lagrangian relaxation.   
% Due to the page limit for the conference
%Before going to the details of  solution (deriving policy) of {{Problem 2}},
   % some definitions are presented below.
% \begin{Defi}(Classes of policies)
% \begin{itemize}
%     \item Stationary Policy: A policy is stationary if decisions only $\bold{s}[t_1]=\bold{s}[t_2]$, then $\bold{a}[t_1]=\bold{a}[t_2]$.
%     %depends on the current states\textcolor{red}{Not a precise definition}.
%     If  decisions are specified  by a probability distribution on the set of possible decisions, called stationary randomized policy, while stationary deterministic policies specify decisions deterministically.
%     \item Feasible Policy: We say a policy is feasible for the problem \eqref{Org_P}, if it satisfies the constraint.
%     \item Optimal Policy: A feasible policy that minimizes {{Problem 2}}  denoted by $\pi^*$, i.e., $J(\pi^*)=\min_{\pi}J(\pi)$.
% \end{itemize}
% \end{Defi}
%\begin{Pro}
% (Restriction to the Stationary Policies): There exists an optimal stationary policy for the {{Problem 2}}.
% \begin{proof}
% We need to verify that there exists a  stationary feasible policy which finite average AoI cost. ...
% \end{proof}
%\end{Pro}
%\vspace{1em}
%\vspace{-3mm}
\subsection{Lagrange Relaxation Method }
%To handle the CMDP problem \ref{}, 
We transform problem \eqref{Org_P} into an unconstrained average cost MDP (or simply MDP) leveraging the Lagrangian relaxation method. 
 By this method, for a given Lagrange  multiplier ${\lambda}\ge0$, the Lagrangian, acting as the objective function for the MDP problem, 
 %as the objective of the MDP
 is given by
 %(constant term $-\lambda \Gamma_{\max}$ and $\bold{s}[0]$ are omitted for the notation simplicity) 
    % \begin{equation} %\label{Lag}
  %    \scriptsize
    %  \tiny
      %\footnotesize
    % \small
% \nonumber
    $   \mathcal{L}(\pi,{{\lambda}};\bold{s}[0])
         %\\&\nonumber
         \triangleq \limsup_{T\rightarrow \infty} \frac{1}{T}
       \left( \Bbb{E}\left[\sum_{t=1}^{T} %\left(w_1\Delta_1[t]+w_2\Delta_2[t]\right)
      % L(\bold{s}[t],\bold{a}[t];\lambda)
       C(\bold{s}[t])+\lambda D(\bold{a}[t])~|~\bold{s}[0]
      % C(\bold{s}[t])
       %\right]
       % +{\lambda}\Bbb{E}\left[\sum_{t=1}^{T} 
     %+\lambda d(\bold{a}[t])
      \right]
      %+\Bbb{E}\left[\sum_{t=1}^{T} 
      % e^{\pi}[t]
       %\left(\mathds{1}_{\{\alpha[t]\neq 0\}}+\mathds{1}_{\{\beta[t]\neq 0\}}\right)
    %   \right]
        \right)$.
   %  \end{equation}
    % where $L(\bold{s}[t],\bold{a}[t];\lambda)\triangleq C(\bold{s}[t])+\lambda D(\bold{a}[t])$ is the immediate (per-slot) cost of the MDP  for the given $\lambda$. 
    %\textcolor{blue}{
   Note that the term $-\lambda\Gamma_{\max}$ is omitted from the Lagrangian because of being constant with respect to the policy for a given $\lambda$.
An    optimal policy  of  the MDP,  for fixed $\lambda$,  denoted by $\pi^*_{\lambda}$ and called \textit{MDP-optimal policy}, is a solution of  the following MDP problem:
%}
    % \\
    %{Problem 2} \textit{(The MDP problem):}
     \begin{align} %\nonumber
     \label{Pro_MDP}
       \underset{\pi_\lambda}{\text{minimize}} ~~~ \mathcal{L}(\pi_\lambda,{{\lambda}};\bold{s}[0]).
     \end{align}
   \indent
By checking the growth condition \cite[Eq. 11.20]{Eitam_CMDP}, 
%and under that condition,
the optimal value of the CMDP problem \eqref{Org_P}, $J^*(\bold{s}[0])$, and the optimal value of the  MDP problem \eqref{Pro_MDP}, denoted 
%which is indicated 
by $\mathcal{L}^{*}({{\lambda}};\bold{s}[0])$, ensures the following:
\begin{align}\label{Eq_jSup}
%\[
J^{*}(\bold{s}[0])=\sup_{\lambda\ge 0}~ \mathcal{L}^{*}({{\lambda};\bold{s}[0]})-\lambda\Gamma_{\max}.
%\]
\end{align}
%\end{Pro}
% \begin{proof}
% By \cite{Eitam_CMDP}, we need to check growth condition such that:   $\forall l\in\Bbb{R}$, the set $G(l)=\{\bold{s}\in\mathcal{S}:~\inf_{\bold{a}} C(\bold{s},\bold{a})<l \}$ is finite. For given $l$, $\sum_i \theta_i+x_i
% +y_i=\inf_{\bold{a}} C(\bold{s},\bold{a})$. Thus, $G(l)$ is finite.
% \end{proof}
%=================
%\begin{proof} (Checking growth condition \cite{Eitam_CMDP})
%Under progress ...
%We should verify the following condition ....
%\end{proof}
%===================
Therefore, the solution of the CMDP problem \eqref{Org_P} can be found by an algorithm that  iteratively  executes the following  two steps: 1) find an optimal policy of the MDP problem \eqref{Pro_MDP} with fixed $\lambda$, i.e., $\pi_{\lambda}^*$,
and 2) update $\lambda$ to a direction that aims to obtain $J^{*}(\bold{s}[0])$ according to \eqref{Eq_jSup}.
 \\\indent
 In the next section, we 
 %utilize the specific relationship
  focus to find an MDP-optimal policy %together with  
 %between the CMDP and MDP problems in Theorem \ref{Th_ES_OP1} and propose an algorithm to
% find MDP-optimal policy and 
 $\pi^*_{\lambda}$ and estimating the optimal Lagrange multiplier.
% multipliers $\{\lambda^*-\xi,\lambda^*+\xi\}$. 
%The elegance and simplicity of the structure of that policy is clear.
%However, it is hard to find that policy in practice. 
%\\\indent
% provide an algorithm that find the optimal policy of the MDP problem, i.e., $\pi^*_{\lambda}$. Then, find the optimal $\lambda$.
% Therefore, one important step is to find the optimal policy of the MDP problem which is provided next.
%\vspace{-0.5em}
\section{Optimal Policy of the CMDP Problem}\label{Sec_Optimal_MDP}
%\vspace{-0.4em}
%As discussed above,  obtaining an policy for the CMDP problem \eqref{Org_P} necessitates finding  an MDP-optimal policy together with the optimal Lagrange multiplier.
In the following, we focus on obtaining an  MDP-optimal policy and estimating the optimal Lagrange multiplier, respectively, in Sec. \ref{Sec_Opt_MDP} and Sec. \ref{Lag_Eta}.
%In this  section, we  turn
%To find $\pi^*_{\lambda}$ as an optimal policy of MDP problem \eqref{Pro_MDP} and  then in Section \ref{Lag_Eta} we will focus to find ${\lambda^*}$.
\subsection{Optimal Policy of the MDP Problem}\label{Sec_Opt_MDP}
%\textcolor{blue}{
We elaborate on the structure of MDP-optimal policy  for the case with error-free links by the following theorem. 
 \begin{theorem}
   \label{Th_Detr} 
   %(Structure of the optimal policy of {Problem 3}) %and  error-free links
 For given $\lambda$ and error-free links, any MDP-optimal policy of problem \eqref{Pro_MDP}
 has
 %and $\pi^{*}_{\lambda_2}$) is 
     a switching-type structure\footnote{
     A switching-type structure means that if a policy takes action $\beta=i$ at state $\bold{s}$,
     then it takes the same action at all states  $\bold{s}+z\bold{e}_{i+4}$, for all $z\in\Bbb{N}$, where 
     $\bold{e}_{i+4}$ is a vector in $\mathbb{B}^{6}$ in which the $(i+4)$-th element is $1$ and the others are $0$, where $\mathbb{B}$ denotes the field of  binary numbers.
     } %\cite{Eyton_Modiano}
    % \footnote{A switching-type structure means that if an optimal policy takes action $\beta=i$ at state $\bold{s}$,
     %then it takes the same action at all states  $\bold{s}+z\bold{e}_{i+4}$, for all $z\in\Bbb{N}$, where 
    % $\bold{e}_{i+4}$ is a vector in $\mathbb{B}^{6}$ in which the $(i+4)$-th element is $1$ and the others are $0$, where $\mathbb{B}$ denotes the field of  binary numbers.} 
     for $\beta$ with respect to $\bold{y}=(y_1,y_2)$.
     %\label{Footnote_} 
     %Further analysis  is  derived to the extended version.
      %\textcolor{blue}{with respect to the $\beta$}.
 \end{theorem}
 \begin{proof}
 The proof will be provided in the extended version.
% %\textcolor{blue}{We can discuss about it.}
 \end{proof}
 Value iteration is a classical method to handle MDP problems, but cannot be applied for the infinite state  spaces. To circumvent  this problem, we use the state truncation method and approximation analysis  \cite[Ch. 16]{Eitam_CMDP},\cite{Sennot_Book}.
\subsubsection{State Truncation and Approximated MDP}
 We truncate $\mathcal{S}$ into a finite  state space $\mathcal{S}^{(N)}$ which is
 %called the truncated state space and  
 parameterized by an integer $N$.
 %$N$, where $N$ is  an integer number.
 To this end,
whenever the AoIs exceed $N$, we set their values to $N$. 
%For example, the dynamic of  AoI at R is rewritten as
% \begin{align}
% %\label{AoI_R1}
% % \nonumber
% %   &  \theta_i[t]=\begin{cases}
% %  %  \thet\bold{a}[t]\mathds{1}_{\{\alpha[t]=1\}}+
% %  %  \theta_i[t]
% %   0, &\text{if}~~ \Lambda_i [t]=1,
% %     \\
% %     \theta_i[t-1]+1, & \text{otherwise}
% %     \end{cases}, %~i\in\{1,2\},
% %\end{align}
% %The evolution of AoI for 
% %$i\in\mathcal{I}$ 
% %at the R can be characterized by:
% %\begin{align}
% %\\&
% \nonumber
% %\label{AoI_R1}
%     \delta_i[t+1]=\begin{cases}
%  %  \thet\bold{a}[t]\mathds{1}_{\{\alpha[t]=1\}}+
%   \min(\theta_i[t]+1,N) &\text{if}~~ \alpha[t]=i,~p[t]=1,
%     \\
%   \min( \delta_i[t]+1,N), & \text{otherwise}
%     \end{cases}, %~i\in\{1,2\},
% %\end{align}
% %The evolution of AoI at the destination can be characterized by
% %\begin{align}
% % \\&\nonumber
% %     \Delta_i[t+1]=\begin{cases}\label{AoI_D1}
% %     \delta_i[t]+1, &\text{if}~~
% %     \beta[t]=i,~q[t]=1,    %    \delta_{i}[t]+1, &\text{if}~~ q[t]=1,~\beta[t]=i,
% %     \\
% %     \Delta_i[t]+1, &\text{otherwise}
% %     \end{cases}. %~i\in\{1,2\}.
% \end{align}
%Accordingly, the relative AoIs are updated as $x_i[t]=\min(\delta_i[t]$
 %By $\mathcal{S}^{(N)}$, we denote the resulting state space which is parameterized by $N$ %and called the truncated state space. Also,
 The corresponding  MDP is called the truncated or approximated MDP.
The state transition probabilities of the truncated MDP,  $\mathcal{P}_{\bold{s}\bold{s}'}^{(N)}(\bold{a})$, are obtained by Eq. \eqref{Eq_TranPro_Unr} under   the following changes. In all equations of Eq. \eqref{Eq_TranPro_Unr}, we replace $\theta_i+1,~x_i,$ and $y_i$,
%each element  of $\bold{s}'_i$ except the cases such that $x_i'=x_i$ and $y_i'=y_i$,
respectively,
%by $N$, when its value exceeds $N$. The mentioned cases are replaced 
by 
$\theta_i+1\rightarrow\big[\theta_i+1\big]_{N}^{+}$,
$x_i\rightarrow \big[x_i+\theta_i+1\big]_{N}^{+}-\big[\theta_i+1\big]_{N}^{+}$ and  $y_i\rightarrow \big[y_i+x_i+\theta_i+1\big]_{N}^{+}-\big[x_i+\theta_i+1\big]_{N}^{+}$, where  $\big[z\big]_N^+$ equals to $z$ if $z\le N$; otherwise,  it equals to $N$.
%that is denoted by $s'$ , by $F(s')$, where $F(s)=\min\{s,N\}$. 
%========== EXAMPLE COMMENTED
%For example, assume that $\bold{s}=(\theta_1,x_1,y_1;\theta_2,x_2,y_2)$ in $\mathcal{S}^{(N)}$ and action $\bold{a}=(0,2)$, then,  $\bold{s}'=(0,\min(N,\theta_1+x_1+1),\min(N,y_1)$ 
%  \begin{align}
%       \mathcal{P}_{\bold{s}\bold{s}'}^{(N)}(\bold{a})= \mathcal{P}_{\bold{s}\bold{s}'}(\bold{a})+\sum_{\bold{r}(\bold{s}')\in\mathcal{S}-\mathcal{S}^{(N)}}  
% \end{align}
%The resulting  MDP is called truncated or approximated MDP.
%, in which, state space and the transition probability are given by $\mathcal{S}^{(N)}$ and $\mathcal{P}_{\bold{s}\bold{s}'}^{(N)}(\bold{a})$, respectively.  
%Clearly, switch-type structure can be  
%It is noteworthy that 
In general, there is no guarantee that the truncated   MDP converges to the original one \cite[p. 276]{Sennot_Book}. By \cite[Theorem 9]{Eyton_Modiano}, for large enough $N$, the truncated MDP convergences to the original MDP.
\subsubsection{Finding Optimal Policy of the Truncated MDP}\label{Sec_POlicy_Des}
Now, the Relative Value Iteration (RVI) %\cite{Sennot_Value_Iter} 
algorithm can be exploited to solve the truncated MDP for a given  $\lambda$ and parameter $N$. In specific, RVI is an iterative algorithm  with the following value iteration equation
%strating with $V_0^(N)(\bold{s})$ a
%updates the following  rule 
with iteration index $n$:
\begin{align}
\nonumber
    V_{n+1}^{(N)}(\bold{s})=\min_{\bold{a}}\left\{ L(\bold{s},\bold{a};\lambda)+\Bbb{E}[h_n^{(N)}(\bold{s}')]\right\}, %-h_n^{(N)}(\mathbf{s}_0),
\end{align}
where $L(\bold{s},\bold{a};\lambda)=C(\bold{s})+\lambda D(\bold{a})$ and $h_n^{(N)}(\bold{s}')\triangleq V_n^{(N)}(\bold{s}')-V_n^{(N)}(\mathbf{s}_\mathrm{ref})$ is the relative value function,
%\cite[p. 266]{Sennot_Value_Iter},
$\mathbf{s}_\mathrm{ref}$ is a reference state, % which is selected as $\mathbf{s}_{\mathrm{ref}}=(0, 0, 1, 2, 1, 2)$,
and $V_0^{(N)}(\mathbf{s}_\mathrm{ref})=0$.
Moreover,  $\Bbb{E}[h_n^{(N)}(\bold{s}')]=\underset{\bold{s}'\in\mathcal{S}^{(N)}}{\sum}\mathcal{P}^{(N)}_{\bold{s}\bold{s}'}(\bold{a})h_n^{(N)}(\bold{s}')$.
The details of RVI are provided in Alg. \ref{A_RVI} (see Steps 3-11), where $\varepsilon$ is a RVI termination criterion. 
%Note that using the structure demonstrated in Theorem \ref{Th_Detr} decreases the complexity of  RVI. 
%\begin{Defi}($\lambda$-optimal policy)\label{Def_opt}
%An optimal policy for the MDP {Problem 3} for the given $\lambda$ is called a $\lambda$-optimal policy.
%and denoted by $\pi_{\lambda}^*$.
%\end{Defi}
%We employ RVI  to find the optimal policy of the MDP problem that is denoted by $\pi_{\lambda}^*$.
% commented in CR V
%\begin{Pro}\label{Th_uni}
\\\indent
      For a given $N$ and $\lambda$, the RVI algorithm 
      give an optimal policy 
      %converges to the optimal policy point of
     for the truncated  MDP after a finite number of iterations. This is because the truncated state MDP is unichain and by \cite[Theorem 8.6.6]{Puterman_Book} the optimality  of RVI is guaranteed.
     %Further, beining unichain implies that 
      %i.e., $\pi^*_{\lambda}$ is the optimal policy.
%\end{Pro}
% \begin{proof}
% %By \cite[Theorem 8.6.6]{Puterman_Book} and \cite[Ch. 4]{Gallager}, it is sufficient that the finite state Markov chain corresponding to the truncated state MDP, induced by every deterministic stationary policy,  has a state which is accessible from any other state. State $\bold{s}=(N,N,0,0,0,0)$ has such property.
% %Detailed analysis is deferred to the extended version. 
% The proof is deferred to the extended version.  
% \end{proof}
%So far, we provide an algorithm that gives $\pi^*_{\lambda}$ as an optimal policy for MDP problem \eqref{Pro_MDP} for a given  $\lambda$. Next, we turn to find  optimal policy of the  CMDP problem \eqref{Org_P}.
\subsection{ Estimating  Optimal Lagrange Multiplier} \label{Lag_Eta}
%======================= NEW 
% Before going to details\footnote{We frequently use some results of \cite{Sennot_1AC}, which are based on the Assumption 1-5 in \cite{Sennot_1AC}, as stated in the proof of Theorem \ref{Th_ES_OP}. One can not use these results directly, without clearly verifying or assuming that the assumptions hold.}, we present the following proposition, that characterizes the optimal policy of {Problem 2}.
% \begin{Pro}\label{Pro_OptCMDP}
%  Assume that there exist  a multiplier $\lambda>0$ and a policy $\pi$, such that  the following conditions hold:
% \\ C1) $\pi$ is $\lambda$-optimal (see Def. \ref{Def_opt}),
% \\ C2)
%  ${D}(\pi;{\lambda},\bold{s}[0])= \Gamma_{\max}$,
% \\ C3) $\mathcal{L}(\pi;{{\lambda}},\bold{s}[0])=J(\pi,\bold{s}[0])+\lambda D(\pi,\bold{s}[0])$.
% \\
%  By \cite[Lemma 3.10]{Sennot_1AC} or \cite[Theorem 4.3]{Ross_Optimal}, the policy $\pi$ is the optimal policy of the CMDP {{Problem 2}}. 
% \end{Pro}
%  By proof of Theorem \ref{Th_uni}, if $\pi$ be stationary, C2-C3 are independent of $\bold{s}[0]$. Moreover, for {any stationary} $\lambda$-optimal policy, C3 holds immediately.
%\subsection{Finding the Optimal Lagrange Multiplier}
%By verifying Assumption 1-5 of \cite[P. 72-73]{Sennot_1AC} in a similar way, provided in \cite[Appendix A]{Elif_ARQ}, we use some main result of \cite{Sennot_1AC}.
 By \cite[Lemma 3.4]{Sennot_1AC}, for $\lambda>0$,  $J(\pi^{*}_{\lambda})$ and $\mathcal{L}(\pi_{\lambda}^*;\lambda)$  ($\bold{s}[0]$ is omitted) are increasing  in $\lambda$ and $\bar{D}(\pi^{*}_{\lambda})$ is decreasing in $\lambda$.
 %---------------------------
%  \footnote{By proof of Proposition \ref{Th_uni}, 
%  %and the fact that $\pi^*_{\lambda}$ is stationary,
%  $J(\pi^{*}_{\lambda})$, $\mathcal{L}(\pi_{\lambda}^*;\lambda)$, and $\bar{D}(\pi^{*}_{\lambda})$ are independent from  an initial state, and for a given $\lambda$, we have the  equality $\mathcal{L}(\pi_{\lambda}^*;\lambda)=J(\pi^{*}_{\lambda})+\bar{D}(\pi^{*}_{\lambda})$. Note that since their values may vary with  the initial state, the equality does not hold for any policy in general \cite[p. 73]{Sennot_1AC}.}
 %--------------------------------------
 Therefore, the optimal Lagrange multiplier $\lambda^*$ is given by
  \begin{align}
  \lambda^*=\inf\{\lambda>0: \bar{D}(\pi^{*}_{\lambda})\le \Gamma_{\max}\}.
  \end{align}
% \begin{Cor}
% \label{Pro_Inc}
% As the result of \cite[Lemma 3.4]{Sennot_1AC}, for $\lambda>\gamma$, we have ${D}(\pi^{*}_{\lambda})\le \Gamma_{\max}$, whereas for $\lambda<\gamma$, we have ${D}(\pi^{*}_{\lambda})> \Gamma_{\max}$. 
% \end{Cor}
As a result, if we find $\pi^*_{\lambda^*}$ such that $\bar{D}(\pi^*_{\lambda^*})=\Gamma_{\max}$, then $\pi^*_{\lambda^*}$ is  an optimal policy for the  CMDP problem \eqref{Org_P}. Exploiting the monotonicity of $\bar{D}(\pi^*_\lambda)$ with respect to $\lambda$, 
%is monotonically decreasing, when $\lambda$ is increasing.
%\footnote{Since large $\lambda$ does not allow more transmission that makes increasing age and decreasing the number of transmissions as the quantity of the constraint.} 
%Thus,
we adopt the bisection search \cite{Age-Energy_Gursoy,Maatuk_CMDP_AoII} to
find $\lambda^*$. Details are stated in Alg. \ref{A_RVI}, where we initialize  $\lambda^{-}=0$ and $\lambda^{+}$ as a large positive real number, and $\zeta$ is a sufficiently small positive real number for the bisection termination criterion. 
%More precisely, $\zeta$ should  be set in order to guarantee that   $\pi^*_{\lambda^+}$ and $\pi^*_{\lambda^-}$ converges.
\\\indent
Importantly, there is no guarantee (roughly, no possibility), even for \textit{any arbitrarily} small  $\zeta$,
%for any setting of bisection termination criteria $\zeta$ in Alg. \ref{A_RVI},
that $\pi^*_{\lambda_{\mathrm{bis}}}$ obtained by Alg. \ref{A_RVI}  would ensure that  $\bar{D}(\pi^*_{\lambda_{\mathrm{bis}}})=\Gamma_{\max}$. This is, $\pi^*_{\lambda_{\mathrm{bis}}}$ (resp. $J(\pi^*_{\lambda_{\mathrm{bis}}})$) is not necessarily an optimal policy (resp. the optimal value) for the  CMDP problem \eqref{Org_P}.
%\textcolor{blue}{
%\textcolor{blue}{
Following \cite{Adam_Estimation}, by exploiting a mixing policy\footnote{In a mixing policy, randomization (between two deterministic policies)  occurs   exactly \textit{once} before starting to operate the system.} 
which is mixture
%by Theorem \ref{Th_ES_OP1}, an optimal policy of the  CMDP problem \eqref{Org_P} 
    %is
      %a randomized policy,  $\pi_{\mathrm{ran}}$,   {or} 
     % a mixing  policy
      %\footnote{%In a randomized policy, the randomization  occurs at each state with a randomization factor $\eta$, while 
     % In a mixing policy, randomization  occurs   exactly {once} before starting to operate the system.
     % } 
       %$\pi_{\mathrm{mix}}$,
      %, with randomization factor $\eta$,
    %  between %two deterministic
     of % policies
      $\pi^*_{\lambda^-}$ (infeasible policy) and $\pi^*_{\lambda^+}$ (feasible policy)
      %such that:  ${D}(\pi^*_{\lambda^-})>\Gamma_{\max}$, and  ${D}(\pi^*_{\lambda^+})\le\Gamma_{\max}$.
  %   The optimal mixing factor, 
     with mixing factor $\eta$,  the value of CMDP problem \eqref{Org_P}, $J_{\mathrm{mix}}$, is given  by
    % he optimal value of the CMDP problem \eqref{Org_P}
 %consequently, the optimal value of the average sum of AoIs at D as the objective, 
 %is obtained by 
 \begin{align}\label{Eq_optimal_Point}
 %\nonumber
  J_{\mathrm{mix}}=\eta J(\pi^*_{\lambda^+})+(1-\eta)J(\pi^*_{\lambda^-}), %\emph{f}
 \end{align}
 where 
\begin{align}\label{Eq_etaRandomized}
%\tiny
%&\nonumber \eta^*{D}(\pi^*_{\lambda_1}) + (1-\eta^*){D}(\pi^*_{\lambda_2})=\Gamma_{\max}\Rightarrow
 %\\&
 \eta=\frac{\Gamma_{\max}-\bar{D}(\pi^*_{\lambda^{-}})}{\bar{D}(\pi^*_{\lambda^{+}})-\bar{D}(\pi^*_{\lambda^{-}})}.
\end{align}
%}
%  Otherwise, to find an optimal policy one stated in Theorem \ref{Th_ES_OP}, the following steps need to be executed. 
%  %\begin{enumarate}
%  \\ $\bullet$ \textbf{Step 1}: Find polices, say $\overline{\pi}^{*}_{\lambda^*}$ and $\underline{\pi}^{*}_{\lambda^*}$, such that the following conditions are met: C'1) $0< \lambda^*<\infty$, C'2)  ${D}(\overline{\pi}^*_{\lambda^*})>\Gamma_{\max}$ and  ${D}(\underline{\pi}^*_{\lambda^*})<\Gamma_{\max}$, and C'3) $\overline{\pi}^{*}_{\lambda^*}$ and $\underline{\pi}^{*}_{\lambda^*}$ differ at most one state.
%   \\ $\bullet$ \textbf{Step 2}:
%     Find randomization factor $\eta^*$ by
%       \begin{align}
% \label{Eq_RSP}
%     \text{Find}~\eta^*~\text{such that}:~~{D}(\pi^*)=\Gamma_{\max}, ~~ s.t:~ 0\le \eta\le 1,
% \end{align}
% where $\pi^{*}$ can be the randomized
% %\footnote{This policy can be defined symbolically as $\pi_{\text{Ran}}=\eta \overline{\pi}^{*}_{\lambda^*}+(1-\eta)\underline{\pi}^*_{\lambda^*}$.} 
% or the mixing policy of $\overline{\pi}^{*}_{\lambda^*},~\underline{\pi}^*_{\lambda^*}$, and $\eta^*$. Thus, $\pi^{*}$ is the optimal for the CMDP {Problem 2}.
% \\\indent
%  Ac
%       an optimal policy for {Problem 2}, which is,
%       randomized   policy \underline{or} mixing  policy
%       between two deterministic policies, $\pi^*_{\lambda^-}$ and $\pi^*_{\lambda^+}$  such that ${A_RVI}. 
 %Note that, $\lambda^-$ and $\lambda^+$ can mathematically be  represented as $\lambda^-=\lambda^*-\zeta/2$ and $\lambda^+=\lambda^*+\zeta/2$. 
 %\textcolor{blue}{
 We remark that although $J_{\mathrm{mix}}$ is less than $J(\pi^*_{\lambda^+})$,  there  is no guarantee that $J_{\mathrm{mix}}$ is the optimal value of the CMDP problem \eqref{Org_P}. 
%\textcolor{blue}{
This is because  there are no guarantees that  $\pi^*_{\lambda^-}$ and $\pi^*_{\lambda^+}$ are   converged policies\footnote{Here, convergence is point-wise, see \cite[Definition 3.6]{Sennot_1AC}.}, i.e., as $\lambda^-$ increases to $\lambda^*$, the corresponding $\pi^*_{\lambda^-}$ does not change (similarly for $\lambda^+$ as it decreases to $\lambda^*$)  \cite{Ross_Optimal, Sennot_1AC}. %More precisely, to construct a mixing optimal policy using $\pi^*_{\lambda^-}$ and $\pi^*_{\lambda^+}$, these polices must be $\lambda^*$ optimal policies \cite{Ross_Optimal, Sennot_1AC}.
However, $J_{\mathrm{mix}}$ serves as a benchmark to evaluate the performance of any policy such as $\pi^{*}_{\lambda^+}$.
% }
 % CR version
%We remark that $J^*$ as the optimal value of average sum AoI at D serves as a benchmark to evaluate the performance of any policy such as $\pi^{*}_{\lambda^+}$.
%Details of computing $J^*$ are provided in Alg. \ref{A_RVI}.
 \\\indent Finally, we note that the mixing policy is not stationary and ergodic.
 %, and has an unappealing sample path property. This is, under such policy, the sample average corresponding to  the policy does not satisfy the constraint.  
 %Since with probability $\eta$ 
 %given in \eqref{Eq_etaRandomized}, 
 %(resp. $1-\eta^*$), the value of the constrain will be $\bar{D}(\pi^*_{\lambda^+})$ (resp. $\bar{D}(\pi^*_{\lambda^-})$).  
% From this point of view, the randomized policy is more desirable.
 % Meanwhile, to design that,  we have still a challenge to find $\eta^*$. 
 %\\\indent 
 Accordingly, one can select $\pi^{*}_{\lambda^+}$ as a feasible solution  for the CMDP problem \eqref{Org_P} which is a stationary deterministic policy and  is more desirable in practice.
 %----------------------------------------------------
 As explained before, policy $\pi^{*}_{\lambda^+}$ is not guaranteed to be an optimal policy for the CMDP problem \eqref{Org_P}, while it gives acceptable performance for small enough $\zeta$, as shown in the numerical analysis.
\begin{algorithm} [h!]
%\tiny
%\small
\scriptsize
\SetAlgoLined
%\KwResult{Write here the result }
 \KwInput{$N,~\zeta,~\varepsilon,~\lambda^{+},
 %= \lambda_{\text{ini}}^{+},
 ~\lambda^{-},~\Gamma_{\max},~\mu_1,~\mu_2,~ p,~q$}  % initialization\;^
 \While { $|\lambda^{+}-\lambda^{-}|\ge\zeta$}{
 
 $\lambda_{\mathrm{bis}}=\frac{\lambda^{+}+\lambda^{-}}{2}$;
 
 For all $ \bold{s}\in\mathcal{S}^{(N)},~\text{set}~V^{(N)}(\bold{s})=0,~h^{(N)}(\bold{s})=0,~h^{(N)}_{\mathrm{old}}(\bold{s})=1$\;
 
 \While{$
 \max_{\bold{s}\in\mathcal{S}^{(N)}}|h^{(N)}(\bold{s})-h^{(N)}_{\mathrm{old}}(\bold{s})|>\varepsilon$}{
 
 \ForEach  {$\bold{s}\in\mathcal{S}^{(N)}$}{
 
%   \eIf {there exist $z$  such that:  $\beta=i$ for $\bold{s}-z\bold{e}_{i+4}$}{
 
%   $\bold{a}^{*}\leftarrow(\alpha^{*},i)$, where
 
% $\alpha^{*}=\arg\min_{\alpha}\{L(\bold{s},\bold{a};\lambda)+\Bbb{E}\{h^{(N)}(\bold{s}'\})\} $\;
%   }
%   {
 $
 \bold{a}^{*}\leftarrow \underset{\bold{a}
 }{\arg\min}~~ %\arg\min_{\bold{a}\in\mathcal{A}_{\bold{s}}}
 L(\bold{s},\bold{a};\lambda_{\mathrm{bis}})+\Bbb{E}\{h^{(N)}(\bold{s}'\} $\;
 %}
 
 $
 V^{(N)}(\bold{s})\leftarrow L(\bold{s},\bold{a}^{*};\lambda_{\mathrm{bis}})+\Bbb{E}\{h^{(N)}(\bold{s}')\}$\; %-h^{(N)}(\mathbf{s}_0)$\; % \% $\bold{s}\rightarrow\bold{s}'|\bold{a}^{*}$ 
 
 %$h^{(N)}_{tmp}(\bold{s})\leftarrow h^{(N)}(\bold{s})$\;
 
  $h^{(N)}_{\mathrm{tmp}}(\bold{s})\leftarrow V^{(N)}(\bold{s})-V^{(N)}(\mathbf{s}_\mathrm{ref})$\; %\% $\bold{s}_0$ is the reference state\;
  
  }
  $h^{(N)}_{\text{old}}(\bold{s})\leftarrow h^{(N)}(\bold{s})$, $h^{(N)}(\bold{s})\leftarrow h^{(N)}_{\mathrm{tmp}}(\bold{s})$\;
  %V^{(N)}(\bold{s})-V^{(N)}(\mathbf{s}_0)$\;
  
 }
%   \eIf{condition}{
%   instructions1\;
%   instructions2\;
%   }{
%   instructions3\;
%   }
%Derive the optimal policy $\pi^*_{\lambda}$ according to Step 6\;

  Compute $\bar{D}(\pi^*_{\lambda_{\mathrm{bis}}})$\; %following $\pi^*_{\lambda}$\;
  
 \eIf {$\bar{D}(\pi^*_{\lambda_{\mathrm{bis}}})\ge\Gamma_{\max}$}{
 
  $ \lambda^{-}\leftarrow\lambda_{\mathrm{bis}}$\;
  
  }
   { $\lambda^{+}\leftarrow \lambda_{\mathrm{bis}}$\;
  
   }
 }
%$\lambda_{\mathrm{bis}}\leftarrow \lambda$
 Compute  $\pi^*_{\lambda^{+}}$ and $\pi^*_{\lambda^{-}}$\;
 
  Compute  $\bar{D}(\pi^*_{\lambda^{+}})$ and $\bar{D}(\pi^*_{\lambda^{-}})$, and then $\eta^*$ by \eqref{Eq_etaRandomized}\;
  
  Compute $J(\pi^*_{\lambda^+})$ and $J(\pi^*_{\lambda^-})$\;
  
  Compute  %value of average sum AoI  at D, i.e.,
  $J_{\mathrm{mix}}=\eta J(\pi^*_{\lambda^+})+(1-\eta)J(\pi^*_{\lambda^-})$\;
 
  \KwOutput{$\pi^*_{\lambda^+},~J_{\mathrm{mix}}$ } % as a value of the CMDP problem \eqref{Org_P} } %\pi^{*}=\eta \pi^*_{\lambda_1}+(1-\eta)\pi^*_{\lambda_2} \text{in Theorem \ref{Th_ES_OP}}} %\lambda_{1},\lambda_{2}, \pi^{*}_{\lambda}}
 
 \caption{\small Policy design  for the CMDP problem \eqref{Org_P} via  RVI and bisection search %to find $\lambda$-optimal 
} \label{A_RVI}
\end{algorithm} 
\section{Numerical Results}\label{Sec_Numerical_Res}
\begin{figure}[t!]
%$ latex template.tex
%$ dvipdf template.dvi
    \centering
    \includegraphics[width=.32\textwidth]{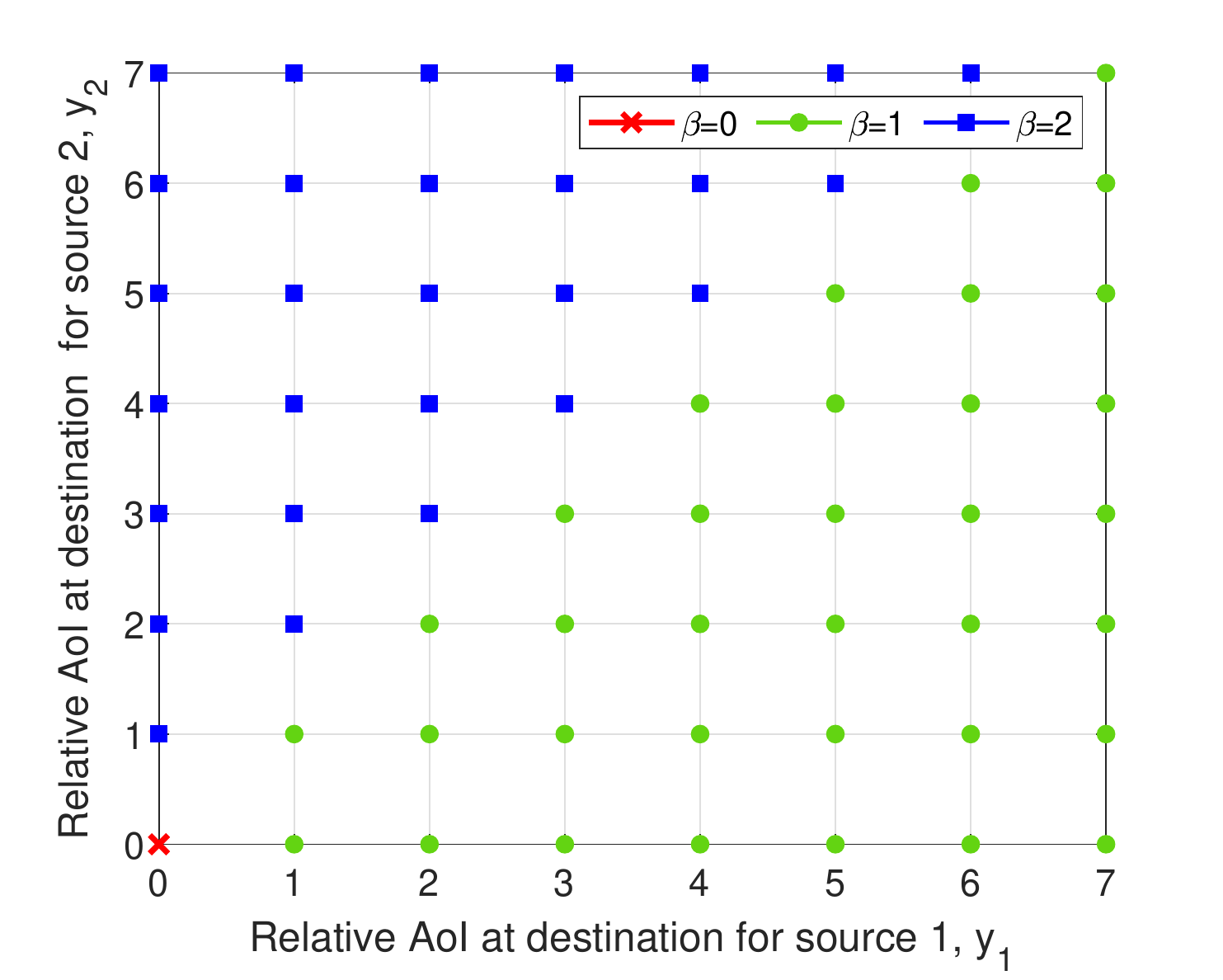}
    \caption{An illustration of the switching-type structure for state $\bold{s}=(1, 2,0,1,y_1,y_2)$ with respect to $\beta$, for $p=0.8,~q=0.7,~\Gamma_{\max}=1.6$, and 
    $\mu=(0.6, 0.9)$.
    \vspace{-5mm}
    }
    \label{Fig_Str_Beta}
\end{figure}

\begin{figure}[t!]
    \centering
    \includegraphics[width=.32\textwidth]{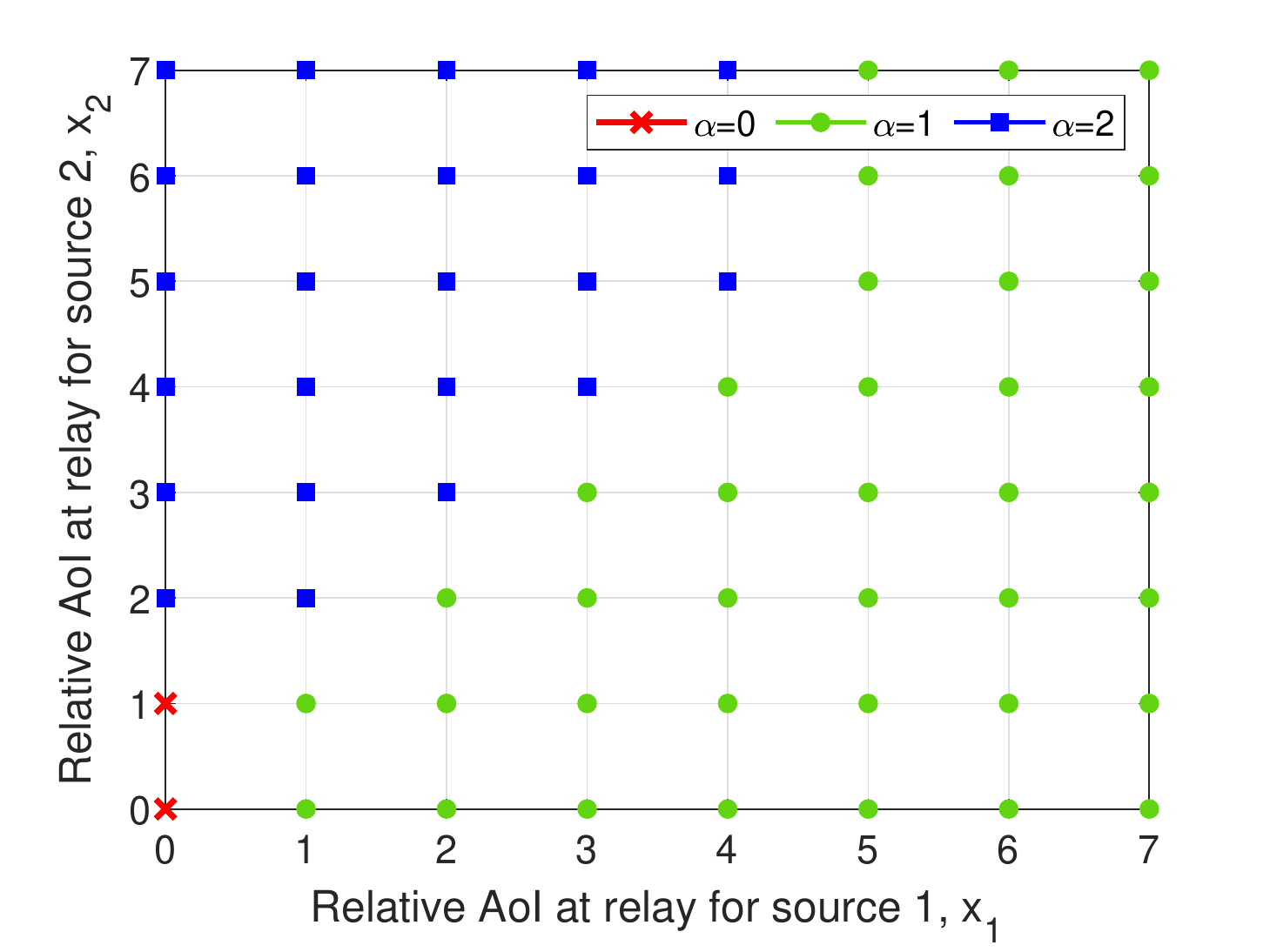}\vspace{-2mm}
    \caption{An illustration of the switching-type structure for state $\bold{s}=(1, 1, x_1,x_2,4,4)$ with respect to  $\alpha$, for $p=0.8,~q=0.7,$ and $\Gamma_{\max}=1.6,~\mu=(0.6, 0.9)$.
       \vspace{-5mm}
    }
    \label{Fig_Str_alpha}
\end{figure}

\begin{figure}[t!]
    \centering
    \includegraphics[width=.5\textwidth]{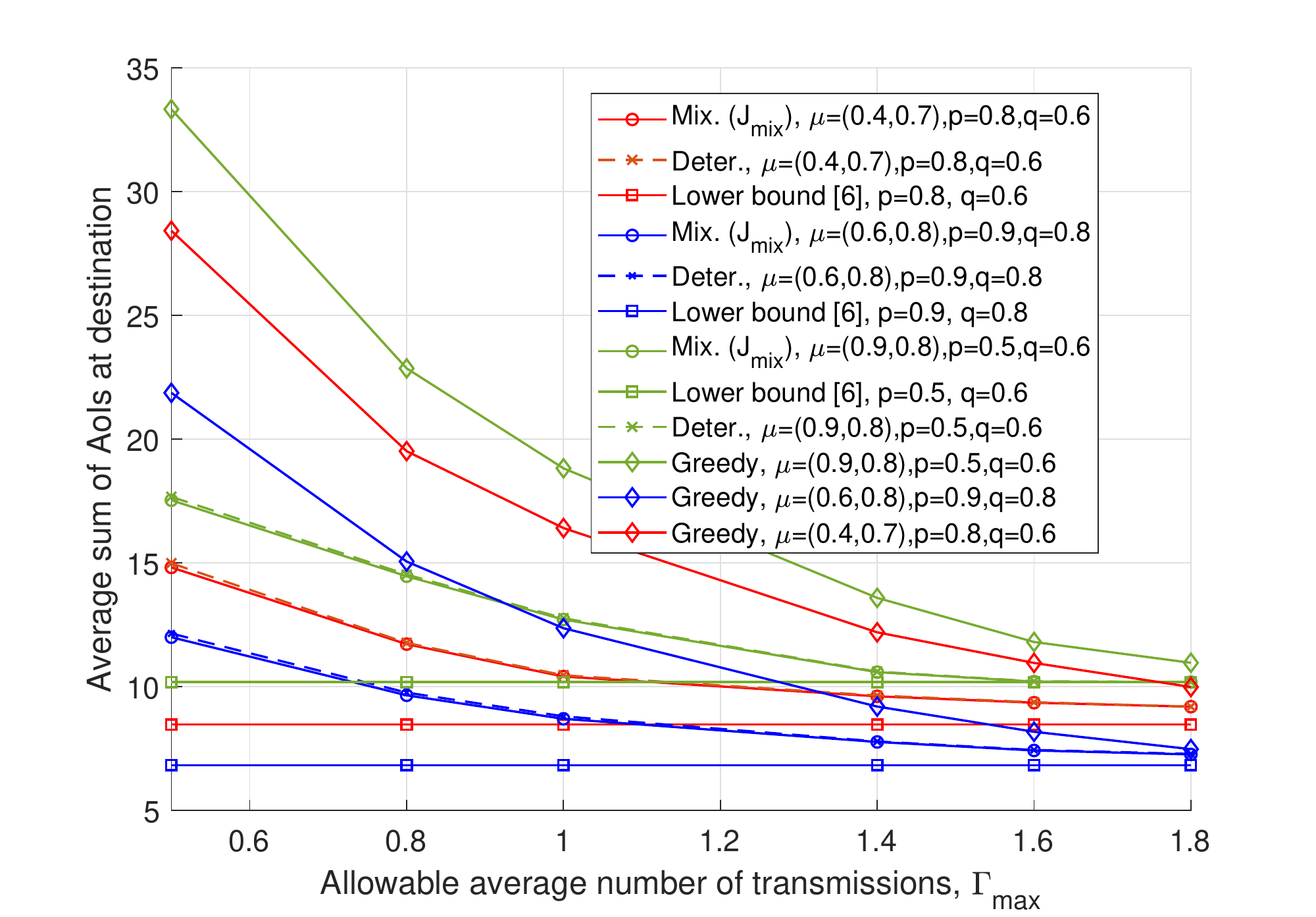}\vspace{-2mm}
    \caption{Average sum AoIs at  D versus $\Gamma_{\max}$, for different arrival rates $\mu=(\mu_1,\mu_2)$, and reliability of channels, $p$  and  $q$. }
    \label{Fig_AAoI_Gaamma}
    \vspace{-5mm}
\end{figure}

In this section, we numerically evaluate  the AoI performance of the proposed algorithm (Alg. \ref{A_RVI}) and the structure of an optimal policy.  The system parameters including the values of the arrival rates, $\mu=(\mu_1,\mu_2)$, the reliabilities of the links in the system, $p$ and $q$, and the allowable average number of transmissions, $\Gamma_{\max}$, are stated  
%The result specific setting, e.g., $\Gamma_{\max},~\mu=(\mu_1,\mu_2),~p$ are 
 in the label or legend of each figure. %The algorithm level parameters are as follows. 
 For Alg. \ref{A_RVI}, we set $N=7$,
 %the bisection and  RVI termination criteria, respectively,   
 $\zeta=0.01$, and $\varepsilon=0.001$. %, and   $N=7$.
\\\indent
%In Fig.s \ref{Fig_Str_Beta}-\ref{Fig_Str_alpha}, we investigate the structure of an optimal deterministic policy of the MDP problem \eqref{Pro_MDP}. 
 Fig. \ref{Fig_Str_Beta} shows an example of the structure of the policy for the  decision at R, $\beta$, with respect to the relative AoIs at D, $y_i$, for state  $\bold{s}=(1, 2,0,1,y_1,y_2)$. This figure verifies Theorem \ref{Th_Detr}, and unveils that
 %when transmission is allowable,
 R schedules the available packet of the  source with a higher relative AoI at D. % when it is scheduled before.
 %For the cases that relative AoIs at D for the sources are close together, R gives a higher priority to the source that has a lower relative AoI at R and update arrival rate; which is the reason for taking action $\beta=1$ at, e.g., ($y_1=1$, $y_2=1$).
 %in Fig. \ref{Fig_Str_Beta}. 
 %This prioritization on source 1 is not very strong, it occurs only in the states where $y_1=y_2$.
%For example, at state $\bold{s}=(1,2,0,1,1,1)$,  packet of source $1$ is sent to D.  
%Since its status update packet conveys the freshest information and hence, sending it more contributes to reducing sum AoIs at D.
\\\indent
Fig. \ref{Fig_Str_alpha} exemplifies the structure of the policy for the  decision at Tx, $\alpha$, with respect to the relative AoIs at R, $x_i$, for state  ${\bold{s}=(1, 1, x_1,x_2,4,4)}$. Having $\alpha=0$ at ($x_1=0$, $x_2=1$) implies that transmission
does not occur
at every state due to the resource budget.
We can conclude from this figure that for fixed $y_1$ and $y_2$, Tx will give a higher priority to schedule the source that has a lower status update rate. This is because  %availability of a new update for that source taking more time in the system, consequently, 
the waiting time at D for receiving new updates from that source becomes large
as a consequence of infrequent packet arrivals,
%taking more time in the system
which becomes a bottleneck in reducing the AoI. % of that source.
\\\indent
Fig. \ref{Fig_AAoI_Gaamma} depicts the average sum AoI at D (sum AAoI) with respect to the allowable average number of transmissions in the system (resource budget), $\Gamma_{\max}$, for different arrival rates and reliability of the links,
%Tx-R link and the R-D link. 
obtained by  averaging over 100,000 time slots.
In this figure, ``Mix." refers to $J_{\mathrm{mix}}$ obtained by  \eqref{Eq_optimal_Point}, and  ``Deter." stands for $J(\pi^*_{\lambda^+})$.
%case in which we exploit the deterministic policy of $\pi^*_{\lambda^+}$ to obtain AAoI.
For benchmarking, we consider a ``Lower bound" scheme where we  eliminate the average resource constraint by setting $\Gamma_{\max}=2$ and realize the \textit{generate-at-will} model by setting the arrival rates as ${\mu_1=\mu_2=1}$.
With these setting, our system becomes equivalent to the one  studied in \cite{Deniz_Relay}, and we use  the greedy-based optimal policy  derived in \cite{Deniz_Relay}. As another benchmark, we design a ``Greedy" policy, where the transmission is allowed in slot $t$ when $\bar{D}_t\le \Gamma_{\max}$, where $\bar{D}_t$ denotes the average number of transmissions until $t$, and the decision criteria are the relative AoI at R for $\alpha$ and the relative AoI at D for $\beta$.
\\\indent
First, Fig. \ref{Fig_AAoI_Gaamma} shows that the deterministic policy, $\pi^*_{\lambda^+}$, achieves a near-optimal performance. Fig. \ref{Fig_AAoI_Gaamma} illustrates that the sum AAoI dramatically increases  when  the resource  budget $\Gamma_{\max}$ is decreased, and  the sum AAoI values become large when channel reliabilities are reduced.
Moreover, Fig. \ref{Fig_AAoI_Gaamma} exhibit that the sum AAoI converges to the lower bound when the arrival rates of updates are high and the resource  budget is large. For that case, we infer that
%Moreover, Fig. \ref{Fig_AAoI_Gaamma}  shows that the average sum AoIs at D
%and values becomes large when channel reliabilities are reduced.
%, and for the small arrival rates, we have large AAoI.
%the impact of channel reliability on performance  is intensive than that of the arrival rates. 
%From this figure, we observe that the deterministic policy achieves a near-optimal performance. 
%However, since the considered metric is an average of AoI in a long-run, i.e., a small performance improvement is more interesting,  designing  optimal policy is pivotal.  
%Finally, according to the structure of the deterministic  policy and Fig. \ref{Fig_AAoI_Gaamma},
%Finally, we  infer from the results when the average resource  budget  is large and status update arrival rates are high,
an optimal policy has a greedy behavior where  the maximum of the relative AoIs at R is the greedy criterion for the decision of Tx, and the maximum of the relative AoIs at D is a greedy criterion for the decision of R. 
%\textcolor{blue}{
We observe that the optimality gap for the Greedy policy is large when the resource budget is small. This clearly emphasizes 
%the importance of an optimal policy design and implies 
that it is pivotal to take into account resource limitations in age-optimal policy design. Namely, it may be possible to find some algorithms that minimize the sum AAoI, but they are not necessarily resource-efficient ones. %Thus, this unveail one motivation of we to deaign an optimal policy under res
%}
%Thus,  one interesting point is that in such cases,  centralized scheduling does not improve the AAoI significantly in compared to  a distributed scheduling\footnote{In the distributed scheduling, the decisions at Tx is based on the AoIs at Tx and R, and the decisions at R is based on the AoIs at R and D. }, as we expected by results of \cite{Deniz_Relay}.
%\textcolor{blue}{Started to implement RVI using bisection method for the Lagrange multiplier}
%Randomized-Greedy as a baseline algorithm
\vspace{-2mm}
\section{Conclusion}
We  studied the sum AAoI minimization scheduling problem in a multi-source relaying system  with stochastic arrivals and  unreliable communication channels, under  per-slot transmission capacity constraints per link and a long-run average resource constraint. To this end, we formulated a stochastic optimization problem and recast it as a CMDP problem. We analyzed the structure of an optimal policy
%, which is in the class of randomized or mixing policies, with a switching-type structure. We 
and proposed an algorithm that obtains  a stationary deterministic near-optimal policy which is easy to implement. 
%We theoretically demonstrated  the existence of the switching-type structure for the case with error-free links. %for the case in which communication links are reliable.
%Moreover, we investigated the effect of system parameters such as arrival rates and the link reliability on the sum AAoI.   
According to the simulation results, our proposed algorithm significantly reduces the sum AAoI compared with the non-trivial greedy-based benchmark algorithm.
We concluded that  when the average resource budget is large and the arrival rates of  status
updates are high, an optimal policy  has a greedy behavior.

\section{Acknowledgments}
This research has been financially supported by the Infotech Oulu, the Academy of Finland (grant 323698), and Academy of Finland 6Genesis Flagship (grant 318927). The work of M. Leinonen has also been financially supported in part by the Academy of Finland (grant 319485).

\vspace{-2mm}
\bibliographystyle{ieeetr}
%{\bibliography{Refbio1}\scriptsize}
%\fontsize{\bibfont}
%\AtBeginBibliography{\small}
\bibliography{Bib/conf_short,Bib/IEEEabrv,Bib/Refbio1}

\begin{thebibliography}{10}

\bibitem{AoI_Mag}
M.~A. {Abd-Elmagid}, N.~{Pappas}, and H.~S. {Dhillon}, ``On the role of age of
  information in the internet of things,'' {\em IEEE Commun. Mag}, vol.~57,
  no.~12, pp.~72--77, Dec. 2019.

\bibitem{EItam_Young}
E.~{Altman}, R.~{El-Azouzi}, D.~S. {Menasche}, and Y.~{Xu}, ``Forever young:
  Aging control for smartphones in hybrid networks,'' CoRR, Sep. 2010.

\bibitem{Roy_2012}
S.~{Kaul}, R.~{Yates}, and M.~{Gruteser}, ``Real-time status: How often should
  one update?,'' in {\em Proc. IEEE Int. Conf. on Computer Commun.},
  pp.~2731--2735, Orlando, FL, USA, Mar. 2012.

\bibitem{MOhammad_1}
M.~{Moltafet}, M.~{Leinonen}, and M.~{Codreanu}, ``On the age of information in
  multi-source queueing models,'' {\em IEEE Trans. Commun.}, vol.~68, no.~8,
  pp.~5003--5017, Aug. 2020.

\bibitem{Marian_Information}
M.~{Costa}, M.~{Codreanu}, and A.~{Ephremides}, ``On the age of information in
  status update systems with packet management,'' {\em IEEE Trans. Inf.
  Theory}, vol.~62, no.~4, pp.~1897--1910, Apr. 2016.

\bibitem{Deniz_Relay}
J.~{Song}, D.~{Gunduz}, and W.~{Choi}, ``Optimal scheduling policy for
  minimizing age of information with a relay,'' {\em arXiv, preprint
  arXiv:2009.02716.}, Sep. 2020.

\bibitem{hatami}
M.~Hatami, M.~Leinonen, and M.~Codreanu, ``{AoI} minimization in status update
  control with energy harvesting sensors,'' {\em arXiv preprint
  arXiv:2009.04224}, Sep. 2020.

\bibitem{Brancu+2Hop}
Y.~{Gu}, Q.~{Wang}, H.~{Chen}, Y.~{Li}, and B.~{Vucetic}, ``Optimizing
  information freshness in two-hop status update systems under a resource
  constraint,'' {\em IEEE J. Sel. Areas Commun.}, vol.~39, no.~5,
  pp.~1380--1392, May, 2021.

\bibitem{Mohammad_DC}
M.~Moltafet, M.~Leinonen, M.~Codreanu, and N.~Pappas, ``Power minimization for
  age of information constrained dynamic control in wireless sensor networks,''
  {\em arXiv preprint arXiv:2007.05364}, Jul. 2020.

\bibitem{MOradi_R}
M.~{Moradian} and A.~{Dadlani}, ``Age of information in scheduled wireless
  relay networks,'' in {\em Proc. IEEE Wireless Commun. and Netw. Conf.
  (WCNC)}, pp.~1--6, Seoul, Korea (South), Mar. 2020.

\bibitem{Relay_Nikoss}
B.~{Li}, H.~{Chen}, N.~{Pappas}, and Y.~{Li}, ``Optimizing information
  freshness in two-way relay networks,'' in {\em Proc. IEEE/CIC Int. Conf. on
  Commun. in China (ICCC)}, pp.~893--898, Chongqing, China, Aug. 2020.

\bibitem{Shroff}
A.~M. {Bedewy}, Y.~{Sun}, and N.~B. {Shroff}, ``Age-optimal information updates
  in multihop networks,'' in {\em Proc. IEEE Int. Symp. Inform. Theory},
  pp.~576--580, Aachen, Germany, Jun. 2017.

\bibitem{2Hop_TWC}
A.~{Arafa} and S.~{Ulukus}, ``Timely updates in energy harvesting two-hop
  networks: Offline and online policies,'' {\em IEEE Trans. Wireless Commun.},
  vol.~18, no.~8, pp.~4017--4030, Aug. 2019.

\bibitem{Relay_SA}
B.~{Li}, Q.~{Wang}, H.~{Chen}, Y.~{Zhou}, and Y.~{Li}, ``Optimizing information
  freshness for cooperative {IoT} systems with stochastic arrivals,'' {\em IEEE
  Internet Things J}, Early Access, 2021.

\bibitem{Vehicule_Book}
Z.~Su, Y.~Hui, T.~H. Luan, Q.~Liu, and R.~Xing, {\em The Next Generation
  Vehicular Networks, Modeling, Algorithm and Applications}.
\newblock Springer, 2020.

\bibitem{Eyton_Modiano}
Y.~P. {Hsu}, E.~{Modiano}, and L.~{Duan}, ``Scheduling algorithms for
  minimizing age of information in wireless broadcast networks with random
  arrivals,'' {\em IEEE Trans. Mobile Comput.}, vol.~19, no.~12,
  pp.~2903--2915, Dec. 2020.

\bibitem{Eitam_CMDP}
E.~Altman, {\em Constrained {M}arkov Decision Processes}.
\newblock volume 7. CRC Press, 1999.

\bibitem{Sennot_Book}
L.~I. Sennott, {\em Stochastic Dynamic Programming and the Control of Queueing
  Systems}.
\newblock volume 504. John Wiley \& Sons, 2009.

\bibitem{Puterman_Book}
M.~L. Puterman, {\em Markov Decision Processes: Discrete Stochastic Dynamic
  Programming}.
\newblock The MIT Press, 1994.

\bibitem{Sennot_1AC}
L.~I. Sennott, ``Constrained average cost {M}arkov decision chains,'' {\em
  Probab. Eng. Inf. Sci.}, vol. 7, no. 1, pp. 69–83, Jan. 1993.

\bibitem{Age-Energy_Gursoy}
H.~{Huang}, D.~{Qiao}, and M.~C. {Gursoy}, ``Age-energy tradeoff in fading
  channels with packet-based transmissions,'' in {\em Proc. IEEE Int. Conf. on
  Computer Commun.}, pp.~323--328, Toronto, ON, Canada, Jul., 2020.

\bibitem{Maatuk_CMDP_AoII}
A.~Maatouk, M.~Assaad, and A.~Ephremides, ``The age of incorrect information:
  an enabler of semantics-empowered communication,'' {\em arXiv preprint
  arXiv:2012.13214}, Dec. 2020.

\bibitem{Adam_Estimation}
D.-J. Ma, M.~Makowski, and A.~Shwartz, ``Estimation and optimal control for
  constrained {M}arkov chains,'' {\em In Proceedings of the 25th IEEE Conf. on
  Decision and Control.}, pp. 994-999, 1987.

\bibitem{Ross_Optimal}
K.~Beutler, F.J.~Ross, ``Optimal policies for controlled {M}arkov chains with a
  constraint,'' {\em Mathematical Methods of Operations Research}, vol. 112,
  pp. 236-252., 1985.

\end{thebibliography}
%\bibliography{Refbio1}{}
%\footnotesize\bibfont
%\sma
\end{document}